\documentstyle[preprint,aps,epsfig]{revtex}
\tightenlines

\begin{document}

\author{F. Giacosa, T. Gutsche and Amand Faessler}
\title{A covariant constituent quark/gluon model for the glueball-quarkonia content
of scalar-isoscalar mesons}
\address{Institut f\"ur Theoretische Physik,
Universit\"at T\"ubingen,\
Auf der Morgenstelle 14,  D-72076 T\"ubingen, Germany }
\maketitle

\begin{abstract}
We analyze the mixing of the scalar glueball with the scalar-isoscalar
quarkonia states above $1$ $GeV$ in a non-local covariant constituent
approach. Similarities and differences to the point-particle Klein-Gordon
limit and to the quantum mechanical case are elaborated. Model predictions
for the two-photon decay rates in the covariant mixing scheme are indicated.
\end{abstract}

\section{\protect\medskip Introduction}

The possible existence and observable nature of the lowest-lying scalar
glueball is currently under intensive debate. The discussion on possible
evidence for the emergence of the glueball ground state in the meson
spectrum has dominantly centered on the scalar-isoscalar resonance $%
f_{0}(1500)$ \cite{close95}. This state has been clearly established by
Crystal Barrel at LEAR \cite{Amsler} in proton-antiproton annihilation, and
is also seen in central pp collisions \cite{bellazzini} and $J/\Psi $ decays 
\cite{bugg}. The main interest in the $f_{0}(1500)$ as a possible candidate
with at least partial glueball content rests on several phenomenological and
theoretical observations. The $f_{0}(1500)$ is produced in gluon rich
production mechanisms, whereas no signal is seen in $\gamma \gamma $
collisions \cite{barate}. Lattice QCD, in the quenched approximation,
predicts the lightest glueball state to be a scalar $(J^{PC}=0^{++})$ with a
mass of $1611\pm 30\pm 160$ MeV \cite{michael12003}. Although the decay
pattern of the $f_{0}(1500)$ into two pseudoscalar mesons might be
compatible with a quarkonium state in the scalar nonet of flavor structure $n%
\bar{n}\equiv {\frac{1}{\sqrt{2}}}(u\bar{u}+d\bar{d})$, the observed, rather
narrow total width of about $\Gamma \approx 110$ MeV is in conflict with
quark model expectations of about $\Gamma \geq 500$ MeV \cite
{close95,gutsche}.

There are mainly two major theoretical schemes, each of them split in a
variety of sub-scenarios, which try to extract and point at the features of
the scalar glueball embedded in the scalar meson spectrum. The first and
original ones \cite{close95,weing} consider the possible nonet of scalar
mesons above $1$ $GeV$ which are overpopulated according to the naive
quark-antiquark picture. Experimentally one identifies a surplus state among
the observed $K_{0}^{*}(1430)$, $a_{0}(1450)$, $f_{0}(1370)$, $f_{0}(1500)$
and $f_{0}(1710)$ resonances. The last three isoscalar $f_{0}$ states are
considered to result from the mixing of two isoscalar $\overline{q}q$ states
and the lowest-lying scalar glueball. A recent phenomenological analysis
concerning the hadronic decays of $f_{0}(1370)$, $f_{0}(1500)$ and $%
f_{0}(1710)$ into pairs of pseudoscalar mesons is consistent with the
minimal three-state mixing scenario \cite{close}. The lower lying resonances 
$f_{0}(980)$ and $a_{0}(980)$ are not taken into account since they
presumably are represented by four-quark configurations with a strong
coupling to virtual $K\bar{K}$ pairs \cite{Close:2002zu}.

In a second scenario \cite{ochs} $f_{0}(980)$ and $a_{0}(980)$ are
interpreted as quark-antiquark states, which together with the $f_{0}(1500)$
and $K_{0}^{*}(1430)$ form a nonet. The $f_{0}(980)$ and $f_{0}(1500)$ are
the isoscalar mesons of the nonet with large mixing as in the pseudoscalar
case. The broad resonances $f_{0}(400-1200)$ (or $\sigma$) and $f_{0}(1370)$
are considered as one state which is interpreted as the light glueball. A
similar approach, which allocates most of the glueball strength in the $%
f_0(1370)$, is considered in Ref. \cite{Anisovich:2001zr}. Some works \cite
{Dosch:2002hc} in QCD sum rules also suggest that the glueball component is
present in the low mass $f_0$ state. However, a recent compilation \cite
{Amsler:2004ps} of theoretical arguments in comparison with experimental
signatures seems to favor the first scenario, where the glueball strength is
centered in the $f_0(1500)$ region.

The aim of the present work is to follow up on the three-state mixing
scenario, where the glueball intrudes in the scalar quark-antiquark spectrum
at the level of 1.5 GeV giving rise to the observed $f_{0}(1370),$ $%
f_{0}(1500)$ and $f_{0}(1710)$ resonances. Up to now many studies in this
direction have been performed (\cite{close95,gutsche,close,klempt} and Refs.
therein), considering a mixing matrix linear in the masses and applying a $%
SO(3)$ rotation to find the orthogonal states, which are then identified
with the three resonances mentioned above. The mixed and unmixed states are
linked by an $SO(3)$ matrix $M$ from which one can read the composition of
the resonances in terms of the ''bare'' unmixed states. In the present work
we develop a framework, where the glueball-quarkonia mixing is described in
a covariant field theoretical context. The present analysis proposes a way
to define the mixing matrix $M$ which allows a comparison with the
Klein-Gordon case and with the quantum mechanical limit. The resulting
matrix $M$ is in general not orthogonal which can be traced to the
non-local, covariant nature of the bound-state systems.

Work along these lines has been performed in the context of effective models 
\cite{jaminon,ebert,volkov}, where the glueball degree of freedom is
introduced by a dilaton \cite{ebert,volkov} or by a scalar field constraint
by the QCD trace anomaly \cite{jaminon}. In the present approach we resort
directly to the elementary gluon fields in a non-local interaction
Lagrangian, where the non-locality allows to regularize the model and to
take into account the ''wave function'' of the gluons confined in the
glueball. We also introduce effective non-local quark-quark interactions,
working in the $SU(3)$ flavour limit, in order to describe the bare, unmixed
quark-antiquark states. Mixing of these configurations is introduced on the
basis of the flavor blindness hypothesis and the glueball dominance, where
mixing is completely driven by the glueball. Another interesting feature of
our study is that we obtain breaking of flavour blindness, when considering
the composite nature of the mesons.

As an elementary input we have to specify the non-perturbative quark and
gluon propagators, which should not contain poles up to $\sim 1.8$ $GeV$,
thus avoiding the on-shell manifestation of quarks and gluons (confinement).
Many considerations and characteristics are independent on the form of the
propagators, as will be shown in the numerical results.

The paper is organized as follows: in Section II we present the details of
the model as based on a non-local covariant Lagrangian formalism. Section
III is devoted to the implications of the model, like mixing properties, the
definition of the mixing matrix $M$ and the comparison with other mixing
schemes. We furthermore specify the two-photon decay rates of the mixed $f_0$
states. In Section IV we present our results and the related discussions.
Finally, Section V contains our conclusions. \newline

\section{The model}

We will consistently employ a relativistic constituent quark/gluon model to
compute the mixing of the scalar quarkonium and glueball states related to
the observed $f_{0}$ resonances above 1 GeV. The model is contained in the
interaction Lagrangian 
\begin{equation}
{\cal L}={\cal L}_{q}+{\cal L}_{g}+{\cal L}_{mix}~,
\end{equation}
where ${\cal L}_{q}$ describes the quark sector, ${\cal L}_{g}$ the scalar
glueball and ${\cal L}_{mix}$ introduces the mixing between these
configurations. In the following we present details for the effective
interaction Lagrangian describing the coupling between scalar mesons and
their constituents.

\subsection{Quarkonia sector}

The quark-quark interaction for the isoscalar $\overline{n}n={\frac{1}{\sqrt{%
2}}}(\bar{u}u+\bar{d}d)$ and $\overline{s}s$ meson states above $1$ $GeV$ is
set up by the isospin symmetric Lagrangian 
\begin{equation}
{\cal L}_{q}=\frac{K_{N}}{2}\left( J_{n}^{2}(x)+J_{s}^{2}(x)\right) ~
\label{lagns}
\end{equation}
where the scalar quark currents are given by the non local expressions: 
\begin{equation}
J_{n}(x)=\frac{1}{\sqrt{2}}\int d^{4}y\left( \overline{n}(x+y/2)n(x-y/2)%
\right) \Phi (y)~\text{with}~n(x)=\left( 
\begin{array}{l}
u(x) \\ 
d(x)
\end{array}
\right)
\end{equation}
and 
\begin{equation}
J_{s}(x)=\int d^{4}y\left( \overline{s}(x+y/2)s(x-y/2)\right) \Phi (y).
\end{equation}
The Lagrangian (\ref{lagns}) is quartic in the quark field as in the $NJL$
model \cite{hatsuda,klevansky,weise}, but with a non-local separable
interaction as in \cite{anikin}. The function $\Phi (y)$ represents the
non-local interaction vertex and is related to the scalar part of the
Bethe-Salpeter amplitude. In momentum space we have $\widetilde{\Phi }(q)$,
the Fourier transform of $\Phi (y)$; the vertex function, which
characterizes the finite size of hadrons, is parameterized by a Gaussian
with 
\begin{equation}
\widetilde{\Phi }(q_{E}^{2})=exp[-q_{E}^{2}/\Lambda ^{2}] \label{fivf},
\end{equation}
where $q_{E}$ is the Euclidean momentum. This particular choice for the
vertex function preserves covariance and was used in previous studies of
light and heavy hadron system \cite{lyubo,ivanov,anikin,rusetsky}. Any
covariant choice for $\widetilde{\Phi }$ is appropriate as long as it falls
off sufficiently fast to render the resulting Feynman diagrams ultraviolet
finite. The size parameter $\Lambda $ will be varied within reasonable
range, checking the dependence of the results on it. 

The Lagrangian (\ref{lagns}) represents the isoscalar sector of the more
general quartic nonlocal $SU_{V}(3)$ flavour-symmetric Lagrangian \cite{anikin}
\begin{equation}
{\cal L}=\frac{K_{N}}{4}\sum_{a=0}^{8}\left( J^{a}\right) ^{2}
\end{equation}
where $J^{a}(x)=\int d^{4}y\overline{q}(x+y/2)\lambda ^{a}q(x-y/2)\Phi
(y^{2})$ with $q^{t}=\left( 
\begin{array}{lll}
u, & d, & s
\end{array}
\right) ,$ $\lambda ^{a}$ are the Gell-Mann matrices with $\lambda ^{0}=%
\sqrt{2/3}1.$ In the isoscalar sector, i.e. considering the $a=0$ and $a=8$ components, we
are left with (\ref{lagns}), which we expressed in terms of the $\sqrt{1/2}(%
\overline{u}u+\overline{d}d)$ and $\overline{s}s$ configurations. In
this work we do not consider the axial $SU_{A}(3)$ transformations, i.e. we do
not relate the scalar quarkonia nonet to the pseudoscalar one. Although
this would be a desirable feature, the connection of these
nonets is not easy to establish. Such an approach with a chirally invariant
interaction is for example pursued in the NJL model, but the predicted
scalar masses are below $1.5$ $GeV$ \cite{hatsuda,klevansky} in conflict with the observed 
experimental resonances.
In the following we restrict the study to the scalar sector,
more precisely on the scalar-isoscalar states, which in turn can mix
with the glueball configuration.

The coupling strength
of the current-current interaction is denoted by $K_{N}$ and is in turn
related to the meson masses. The scalar meson masses $M_{N}$ and $M_{S}$ are
deduced from the poles of the T-matrix and are given by the solutions of the
Bethe-Salpeter equations: 
\begin{equation}
K_{N}-\frac{1}{\Sigma _{N}(M_{N}^{2})}=K_{N}-\frac{1}{\Sigma _{S}(M_{S}^{2})}%
=0~.  \label{mnms}
\end{equation}
The mass operator $\Sigma _{N(S)}(p^{2}),$ where $p$ is the meson momentum,
is deduced from the quark loop diagram of Fig. 1 and given by 
\begin{equation}
\Sigma _{N(S)}(p^{2})=-iN_{c}\int \frac{d^{4}q}{(2\pi )^{4}}Tr\left[
S_{n(s)}(q+p/2)S_{n(s)}(q-p/2)\right] \widetilde{\Phi }^{2}(q^{2});
\end{equation}
where $N_{c}=3$ is the number of colors and $S_{n(s)}$ is the quark
propagator. A Wick rotation is then applied in order to calculate $\Sigma
_{N(S)}(p^{2}).$ As evident from (\ref{mnms}), $M_{N}$ and $M_{S}$ are not
independent. Once the size parameter $\Lambda $ is chosen and using the bare
nonstrange meson mass $M_{N}$ as an additional input, the coupling constant $%
K_{N}$ and the bare strange meson mass $M_{S}$ are fixed.

We now turn to the discussion of the quark propagator. The general form \cite
{alkoferrev} is 
\begin{equation}
S(p)=i\frac{Z(p^{2})}{\left( p\!\!\!\slash-m(p^{2})\right) }.
\end{equation}
Considering the low energy limit we have $Z(p^{2})\sim $constant and $%
m(p^{2})\sim m^{*}$, where $m^{*}$ is the effective or constituent quark
mass. These masses are typically in the range of $0.25$ to $0.45$ $GeV$ for
the $u(d)$ quarks and $0.5$ to $0.7$ GeV for the $s$ flavour. The low energy
limit for the quark propagator is too naive for our purposes, since it leads
to poles in the mass operator for meson masses of about $1.5$ $GeV$. In the
following we consider two possible ways to avoid these infinities. The mass
function for $n=u,d$ flavor can be decomposed as $m_{n}(p^{2})=m_{n}^{*}+%
\sigma _{n}(p^{2})$ and the contribution of $\sigma _{n}$ is assumed to be
replaced by a large average value. To avoid poles, that is the unphysical
decay of the quark-antiquark meson into two quarks, the averaged mass
function is chosen as $\left\langle m_{n}(p^{2})\right\rangle
=m_{n}^{*}+\left\langle \sigma _{n}(p^{2})\right\rangle =\mu _{n}\geq 0.86$ $%
GeV$ (here for the u- and d -flavour), where $2\cdot \mu _{n}$ is a lower
limit set by the mass of the $f_{0}(1710)$ resonance. We then have 
\begin{equation}
S_{n}(p)=\frac{i}{\left( p\!\!\!\slash-\mu _{n}\right) }.
\end{equation}
We also introduce an analogous parameter $\mu _{s}$ for the $s$ flavour,
which for $\mu _{s}\neq \mu _{n}$ contains flavor symmetry violation. Above
parametrization of the quark propagator is the simplest choice; although it
neglects the $p^{2}$ dependence of the quark mass, the approximation by a
free quark propagator with a large effective mass allows to test the
approach and leads to considerable simplifications concerning the technical
evaluation. A similar quark propagator was also used in Ref. \cite{jaminon},
where the decay of scalar mesons into two photons was analyzed.

We also consider a quark propagator which is described by an entire
function: 
\begin{equation}
S_{i}(p)=\frac{i}{p\!\!\!\slash-m_{i}}\left( 1-exp\left( \beta \left(
p^{2}-m_{i}^{2}\right) \right) \right) .  \label{propent}
\end{equation}
This parametrization has been used both in the study of meson and baryon
properties \cite{reinhart,oettel} and serves as one possible way to model
confinement; the factor multiplying the free quark propagator removes the
pole and on-shell $\overline{q}q$ creation is avoided. The effective quark
masses $m_{i}$ (with $i=n,s$) are taken from \cite{alkofer} with $%
m_{n}=0.462 $ $GeV,$ $m_{s}=0.657$ $GeV.$ In Ref. \cite{alkofer} the
effective quark masses are calculated within a Dyson-Schwinger approach and,
as functions of the Euclidean momentum, display an almost constant behavior
up to high values. The parameter $\beta $ is constrained from below to
generate a behavior like $i/\left( p\!\!\!\slash-m_{i}\right) $ for small
(and for euclidean) momenta; $\beta $ is also constrained from above,
requiring that the propagator does not diverge for momenta up to $\sim 2$ $%
GeV$.

The preceding discussion concerning the set up of the quark interaction
Lagrangian ($\ref{lagns}$) and the resulting generation of quark-antiquark
meson states can also be alternatively described. By introducing the
auxiliary meson fields $N(x)$ and $S(x)$ ($N$ for the $\sqrt{\frac{1}{2}}(%
\overline{u}u+\overline{d}d)$ and $S$ for the $\overline{s}s$ meson)
previous procedure can be summarized by the Lagrangian \cite{klevansky}: 
\begin{equation}
{\cal L}_{q}^{\prime }=g_{N}N(x)J_{n}(x)+g_{S}S(x)J_{s}(x)
\label{lagnsprime}
\end{equation}
with the condition that 
\begin{equation}
g_{N(S)}=\left( \frac{\partial \Sigma _{N(S)}(p^{2})}{\partial p^{2}}\right)
_{p^{2}=M_{N(S)}^{2}}^{-1/2}.  \label{compcond}
\end{equation}
The last relation is known as the compositeness condition, originally
discussed in \cite{history} and extensively used in the study of hadron
properties \cite{efimov}). The compositeness condition requires that the
renormalization constant of the meson fields $N(x)$ and $S(x)$ is set to
zero, hence physical meson states are exclusively described by the dressing
with the constituent degrees of freedom. ${\cal L}_{q}^{\prime }$ contains
the meson quark-antiquark vertex with a momentum-dependent, non-local vertex
function. The resulting mass operator is completely analogous to the bubble
diagram of Fig. 1. The Lagrangian (\ref{lagnsprime}) is very useful to
calculate the decays of the bound state $N(S);$ the coupling constant $%
g_{N(S)}$ directly enters in the decay rate.

\subsection{Glueball sector}

We now extend the model to include the scalar glueball degree of freedom, 
described as a bound state of two constituent gluons. 
First we consider the scalar gluonic current $J=F_{\mu \nu }^{a}(x)F^{a,\mu
\nu }(y)$ where $F_{\mu \nu }^{a}(x)=\partial _{\mu }A_{\nu
}^{a}(x)-\partial _{\nu }A_{\mu }^{a}(x)+gf_{abc}A_{\mu }^{b}A_{\nu }^{c}$
and $A_{\mu }^{a}(x)$ is the gluon field. To set up a scalar glueball state
we restrict the current to its minimal configuration, such that the glueball
is described by a bound state of two constituent gluons. With this truncation
we have $\widetilde{J}=\widetilde{F}_{\mu \nu }^{a}(x)\widetilde{F}^{a,\mu
\nu }(y)$ where $\widetilde{F}_{\mu \nu }^{a}(x)=\partial _{\mu }A_{\nu
}^{a}(x)-\partial _{\nu }A_{\mu }^{a}(x)$ is the abelian part of the gluonic
field tensor. The effective, non-local interaction Lagrangian for the constituent gluons is
then written as: 
\begin{equation}
{\cal L}_{g}=\frac{K_{G}}{2}J_{g}^{2}(x),  \label{lagg}
\end{equation}
where 
\begin{equation}
J_{g}(x)=\int d^{4}y\left( \widetilde{F}_{\mu \nu }^{a}(x+y/2)\widetilde{F}%
^{a,\mu \nu }(x-y/2)\right) \Phi (y).
\end{equation}
The non-perturbative features of the gluon dynamics are assumed to be taken
into account by the coupling constant $K_{G}$. It should be stressed that the gluons introduced 
in this section are not the "background" gluons responsible for confinement 
\cite{simonov,simonovlect},
but two effective-constituent gluons forming the glueball. 
 For simplicity we use the
same vertex function as in the previous case, setting the glueball size
equal to the one of the quarkonia states. The coupling constant $K_{G}$
again is linked to the bare glueball mass by the pole-equation: 
\begin{equation}
K_{G}-\frac{1}{\Sigma _{G}(M_{G}^{2})}=0,  \label{kg}
\end{equation}
where the mass operator $\Sigma _{G}(p^{2})$, as indicated in Fig. 2, is
given as: 
\begin{equation}
\Sigma _{G}(p^{2})=i2(N_{c}^{2}-1)\int \frac{d^{4}q}{(2\pi )^{4}}\left[
8(q_{1}\cdot q_{2})^{2}+4q_{1}^{2}q_{2}^{2}\right] D(q_{1}^{2})D(q_{2}^{2})%
\widetilde{\Phi }^{2}(q^{2}).  \label{eq}
\end{equation}
where $q_{1}=q+p/2$ and $q_{2}=-q+p/2.$ $D(q^{2})$ is the scalar part of the
gluon propagator, which in the Landau gauge is \cite{mandula} 
\begin{equation}
D_{\mu \nu }^{ab}=\delta ^{ab}(g_{\mu \nu }-\frac{q_{\mu }q_{\nu }}{q^{2}}%
)D(q^{2}).
\end{equation}

As before, the formalism can alternatively be defined by a Lagrangian
containing the scalar glueball field $G(x)$ with: 
\begin{equation}
{\cal L}_{g}^{\prime }=g_{G}G(x)J_{g}(x),
\end{equation}
where the coupling $g_{G}$ is deduced from the compositeness condition 
\begin{equation}
g_{G}=\left( \frac{\partial \Sigma _{G}(p^{2})}{\partial p^{2}}\right)
_{p^{2}=M_{G}^{2}}^{-1/2}.  \label{compcondglue}
\end{equation}
The last two defining equations are equal to the ones used in \cite{muradov}.

For the gluon propagator we choose the free one \cite{mandula,sorella} 
\begin{equation}
D(q^{2})=\frac{i}{q^{2}-m_{g}^{2}},\label{gluonprop}
\end{equation}
where the effective mass $m_{g}$ should be large, with $2\cdot m_{g}$ larger
than the bare glueball mass deduced from lattice simulations between 1.4-1.8
GeV. \cite{michael12003,weing2,bali,morningstar}. In the Wilson loop approach
of Refs {\cite{simonovlect,simonovmass} a similar scalar 
glueball mass, around 1.58 GeV, is found. 

The generation of a
constituent gluon mass is analogous to the effective quark mass, and one can
relate its value to the gluon condensate \cite{natale}. Typical values for
the effective gluon mass are in the range of 0.6-1.2 $GeV$. Lattice
simulations give a value of 0.6-0.7 $GeV$ \cite{langfeld}; in the QCD
formulation of \cite{kondo} a higher value of about 1.2 GeV is deduced. A
similar value is also found in \cite{amemya} for the off-shell gluon mass in
the maximal abelian gauge. The gauge (in)dependence of the
effective gluon mass is still an open issue \cite{sorella,sorellatalk}. For
our purposes it is sufficient to take a constituent gluon mass within the
range of 0.6-1.2 GeV; we choose an intermediate value of 0.9 GeV, which is in
accord with the effective constituent gluon mass found in the study of
gluon-dynamics in \cite{gubankova}.
In the quarkonia sector we use a quartic interaction Lagrangian; this means
that the background gluons, responsible for the string tension, i.e. for the
attraction among the $\overline{q}-q$ pair, are described by a constant
propagator. The gluons, exchanged among the quark-antiquark pair, are
''soft''; the gluon propagator of eq (\ref{gluonprop}) becomes a constant for small
momenta, thus being in accord with this assumption. The exchange of
non-perturbative gluons is also responsible for the appearance of a vertex
function, i.e. for the finite size of the quarkonia mesons.
In the current model calculation the vertex function
is simply parametrized by (\ref{fivf}).

These arguments are also valid for the soft background gluons exchanged by
the two constituent gluons forming the glueball. On the other hand, when
evaluating the glueball mass-operator (Eq. (\ref{eq}) and Fig. 2), the two
constituent gluons can also have large timelike momenta because of the large
glueball mass ($\sim 1.5$ $GeV)$. In this case the use
of a constant for the gluon propagator is not adequate. This is why we use
the form of Eq (\ref{gluonprop}) with an effective mass.

\subsection{Mixing sector}

We also have to introduce a term which generates the
mixing between the bare glueball and the quarkonia states. Using the flavour
blindness hypothesis, which states that the scalar gluonic current only
couples to the flavor singlet quark-antiquark combination, we have: 
\begin{equation}
{\cal L}_{mix}=K_{SG}J_{g}(x)\left( J_{s}(x)+\sqrt{2}J_{n}(x)\right) ,
\label{lagmix}
\end{equation}
where no direct coupling between the scalar quark currents $J_{n}$ and $%
J_{s} $ has been taken into account. In Eq. (\ref{lagmix}) the two constituent gluons
forming the glueball can transform into a
constituent quark-antiquark couple forming the quarkonia meson. It should be stressed 
that Eq. (\ref{lagmix}) does not describe the fundamental quark-gluon interaction, 
but an effective coupling of the scalar glueball current with the scalar quarkonia ones,
suitable for the description of the mixing among these configurations.

Following the arguments of \cite
{close95} a direct mixing of the bare quarkonia states is a higher order
perturbation in the strong coupling eigenstates and is neglected in the
following.

Due to the mixing term physical states are linear combinations of the bare
quark-antiquark and glueball states. Implications of this scenario will be
studied in the next section.

\section{\protect\medskip Model Implications}

\subsection{ Masses of the mixed states}

We now consider the total Lagrangian containing the interaction terms of (%
\ref{lagns}), (\ref{kg}) and (\ref{lagmix}) with 
\begin{equation}
{\cal L}={\cal L}_{q}+{\cal L}_{g}+{\cal L}_{mix}.
\end{equation}
Due to the mixing term ${\cal L}_{mix}$ the Bethe-Salpeter equations (\ref
{mnms}) and (\ref{kg}) are not valid anymore. In fact, the T-matrix is
modified by ${\cal L}_{mix}$ and takes the form: 
\begin{equation}
T=-(1-K\cdot \Sigma )^{-1}K=-(K^{-1}-\Sigma )^{-1},  \label{t}
\end{equation}
with the non-diagonal coupling matrix 
\begin{equation}
K=\left( 
\begin{array}{lll}
K_{N} & \sqrt{2}K_{SG} & 0 \\ 
\sqrt{2}K_{SG} & K_{G} & K_{SG} \\ 
0 & K_{SG} & K_{S}=K_{N}
\end{array}
\right) ,
\end{equation}
and the diagonal mass operator 
\begin{equation}
\Sigma (p^{2})=\left( 
\begin{array}{lll}
\Sigma _{N}(p^{2}) & 0 & 0 \\ 
0 & \Sigma _{G}(p^{2}) & 0 \\ 
0 & 0 & \Sigma _{S}(p^{2})
\end{array}
\right) .
\end{equation}

The masses of the mixed states are obtained from the zeros of the
determinant in the denominator of the $T$ matrix with $Det\left[ 1-K\Sigma
\right] $\medskip $=0$. In the limiting case of $K_{SG}=0$ the determinant
reduces to $(1-K_{N}\Sigma _{N}(p^{2}))(1-K_{S}\Sigma
_{S}(p^{2}))(1-K_{G}\Sigma _{G}(p^{2}))=0$, which results in the defining
equations of the bare masses without mixing (eqs. (\ref{mnms}) and (\ref{kg}%
)).

For $K_{SG}\neq 0$ we have mixing, where, starting from the unrotated masses 
$\left( M_{N},M_{G},M_{S}\right) ,$ we end up with the mixed states $%
(N^{\prime },G^{\prime },S^{\prime })$ of masses $(M_{N^{\prime
}},M_{G^{\prime }},M_{S^{\prime }})$, which we interpret as the physical
resonances $f_{0}(1370)$, $f_{0}(1500)$ and $f_{0}(1710)$. Quantitative
predictions in the three-state mixing schemes strongly depend on the assumed
level ordering of the bare states before mixing. In previous works \cite
{close95,weing} essentially two schemes were considered: $M_{N}<M_{G}<M_{S}$
and $M_{N}<M_{S}<M_{G}$. In phenomenological studies latter level ordering
seems to be excluded, when analyzing the hadronic two-body decay modes of
the $f_{0}$ states \cite{gutsche,close}. In the current work we will refer
to the first of these two possibilities, where the bare glueball mass is
centered between the quarkonia states before mixing. We also compare our
pole equation $Det\left[ 1-K\Sigma \right] $\medskip $=0$ to other
approaches. But first we discuss further consequences of the covariant
non-local approach, such as the meson-constituent coupling constants, which
are crucial in the calculation of the decays of the mixed states.

\subsection{Resulting meson-constituent coupling constants}

As a result of the mixing the rotated fields $N^{\prime },$ $G^{\prime
},S^{\prime }$ couple to the $\overline{n}n,$ $gg$ and $\overline{s}s$
configurations. The leading contribution to the $T$-matrix (\ref{t}) in the
limit $p^{2}\simeq M_{i}^{2}$ with $i=N^{\prime },G^{\prime },S^{\prime }$
is given by the pole at $p^{2}=M_{i}^{2}$ with 
\begin{equation}
T^{ab}=\frac{g_{i}^{a}g_{i}^{b}}{p^{2}-M_{i}^{2}},
\end{equation}
where $a,b=\overline{n}n,$ $gg$ and $\overline{s}s$ refer to the constituent
components. A similar expression for the $T$ matrix is also given in \cite
{hatsuda,klevansky}, where the $\eta -\eta ^{\prime }$ mixing was analyzed
in the context of the $NJL$ model. The constant $g_{G^{\prime }}^{\overline{n%
}n}$, for example, represents the coupling of the mixed state $G^{\prime }$
to the quark-antiquark configuration of flavor $u$ and $d$. The modulus of
the nine coupling constants is then given by

\begin{equation}
\left| g_{i}^{a}\right| =\lim_{p^{2}\rightarrow M_{i}^{2}}\sqrt{%
(p^{2}-M_{i}^{2})T^{a,a}}=\left( \frac{\partial (T^{a,a})^{-1}}{\partial
p^{2}}\right) _{p^{2}=M_{i}^{2}}^{-1/2}  \label{couplingmod}
\end{equation}
which we solve numerically. In Appendix A, where we consider the mixing of
two fields , we also give an explicit expression for this quantity. The
expression (\ref{couplingmod}) is the generalization of the compositeness
condition (eqs. (\ref{compcond}) and (\ref{compcondglue})).

In the limit $K_{SG}\rightarrow 0$ the T-matrix is diagonal and the coupling
constants $g_{G^{\prime }}^{a}$, for instance, become 
\begin{eqnarray}
g_{G^{\prime }}^{\overline{n}n} &=&g_{G^{\prime }}^{\overline{s}s}=0,\text{ }
\\
\left| g_{G^{\prime }}^{gg}\right| &=&g_{G}^{gg}=\left( \frac{\partial
\Sigma _{G}(p^{2})}{\partial p^{2}}\right) _{p^{2}=M_{G^{\prime
}}^{2}=M_{G}^{2}}^{-1/2}
\end{eqnarray}
where the last equation is the glueball compositeness condition (\ref
{compcondglue}). In this case there is no mixing and the glueball couples
only to gluons.

We still have to discuss the sign of the coupling constants. By convention
we chose $g_{N^{\prime }}^{\overline{n}n},$ $g_{G^{\prime }}^{gg}$ and $%
g_{S^{\prime }}^{\overline{s}s}$ as positive numbers. The sign is determined
from the off-diagonal elements of the T-matrix: 
\begin{equation}
g_{i}^{a}=sign(\alpha _{i}^{a})\left( \frac{\partial (T^{a,a})^{-1}}{%
\partial p^{2}}\right) _{p^{2}=M_{i}^{2}}^{-1/2}  \label{pippo}
\end{equation}
where 
\begin{equation}
\alpha _{i}^{a}=\lim_{p^{2}\rightarrow
M_{i}^{2}}(p^{2}-M_{i}^{2})T^{i,a}=\left( \frac{\partial (T^{i,a})^{-1}}{%
\partial p^{2}}\right) _{p^{2}=M_{i}^{2}}^{-1}.  \label{alfa}
\end{equation}

Having obtained these coupling constants we again can write an effective
interaction Lagrangian for the $G^{\prime }$ field

\begin{equation}
{\cal L}_{G^{\prime }}^{\prime }=g_{G^{\prime }}^{\overline{n}n}G^{\prime
}J_{n}+g_{G^{\prime }}^{gg}G^{\prime }J_{g}+g_{G^{\prime }}^{\overline{s}%
s}G^{\prime }J_{s},  \label{lagglueprime}
\end{equation}
where the coupling between the mixed states and the constituent
configurations are made explicit. This Lagrangian allows to calculate the
decay of $G^{\prime }$ as we will see explicitly for the two-photon decay;
the coupling constants directly enter in the decay rate of the state. One
then has completely analogous expressions for the fields $N^{\prime }$ and $%
S^{\prime }$: 
\begin{eqnarray}
{\cal L}_{N^{\prime }}^{\prime } &=&g_{N^{\prime }}^{\overline{n}n}N^{\prime
}J_{n}+g_{N^{\prime }}^{gg}N^{\prime }J_{g}+g_{N^{\prime }}^{\overline{s}%
s}N^{\prime }J_{s},  \label{lagnprime} \\
{\cal L}_{S^{\prime }}^{\prime } &=&g_{S^{\prime }}^{\overline{n}n}S^{\prime
}J_{n}+g_{S^{\prime }}^{gg}S^{\prime }J_{g}+g_{S^{\prime }}^{\overline{s}%
s}S^{\prime }J_{s}.  \label{lagsprime}
\end{eqnarray}

\medskip The strength of the coupling is directly connected to the mixing
strength, the explicit mixing matrix is discussed in the following.

\subsection{Mixing matrix $M$}

When dealing with elementary scalar particles, and not with composite ones,
mixing between the bare meson fields $(N,G,S)$ can be expressed by the
Lagrangian 
\begin{eqnarray}
{\cal L}_{K-G} &=&\frac{1}{2}(\partial _{\mu }N)^{2}-\frac{1}{2}%
M_{N}^{2}N^{2}+\frac{1}{2}(\partial _{\mu }G)^{2}-\frac{1}{2}M_{G}^{2}G^{2}+%
\frac{1}{2}(\partial _{\mu }S)^{2}-\frac{1}{2}M_{S}^{2}S^{2}+  \nonumber \\
&&+fGS+\sqrt{2}frGN,
\end{eqnarray}
where $f$ and $r$ are mixing parameters ($r\neq 1$ takes into account
breaking of the flavour blindness hypothesis). As before, no direct mixing
between $N$ and $S$ has been taken into account. In this simple case one has
to diagonalize 
\begin{equation}
T_{K-G}=\left( 
\begin{array}{lll}
-M_{N}^{2} & \sqrt{2}fr & 0 \\ 
\sqrt{2}fr & -M_{G}^{2} & f \\ 
0 & f & -M_{S}^{2}
\end{array}
\right)  \label{tkg}
\end{equation}
where the mixing matrix $M_{K-G}$ $\subset SO(3)$ is obtained from the
condition 
\begin{equation}
M_{K-G}\cdot T_{K-G}\cdot M_{K-G}^{t}=D_{K-G}=Diag[-M_{N^{\prime
}}^{2},-M_{G^{\prime }}^{2},-M_{S^{\prime }}^{2}],
\end{equation}
where the primed Klein-Gordon masses refer to the mixed states. The matrix $%
M_{K-G}$ $=M_{K-G}^{i,a}\subset SO(3)$ ($i=N^{\prime },G^{\prime },S^{\prime
}$ and $a=N,G,S)$ connects the unmixed and the mixed states by 
\begin{equation}
\left( 
\begin{array}{l}
\left| N^{\prime }\right\rangle \\ 
\left| G^{\prime }\right\rangle \\ 
\left| S^{\prime }\right\rangle
\end{array}
\right) =M_{K-G}\left( 
\begin{array}{l}
\left| N\right\rangle \\ 
\left| G\right\rangle \\ 
\left| S\right\rangle
\end{array}
\right) .
\end{equation}

In the following we want to determine an analogous matrix in our
Bethe-Salpeter approach. We consider $\left| G^{\prime }\right\rangle $
which can be written as a superposition of the bare states: 
\begin{equation}
\left| G^{\prime }\right\rangle =M^{G^{\prime },N}\left| N\right\rangle
+M^{G^{\prime },G}\left| G\right\rangle +M^{G^{\prime },S}\left|
S\right\rangle
\end{equation}
where $M^{G^{\prime },N}$, for example, is the admixture of $\left|
N\right\rangle =\sqrt{\frac{1}{2}}\left| \overline{u}u+\overline{d}%
d\right\rangle $ to the mixed state $\left| G^{\prime }\right\rangle $.
Unlike in the Klein-Gordon case care should be taken with such an
expression, since here the bare states $\left| N\right\rangle ,\left|
G\right\rangle $ and $\left| S\right\rangle $ are not well defined. They
correspond to the Bethe-Salpeter solutions in the case of zero mixing, but,
when including mixing, they are not normalized vectors of the Hilbert space
anymore. To obtain a corresponding expression as in the Klein-Gordon case,
we proceed as follows: the coupling constant $g_{G^{\prime }}^{\overline{n}%
n} $ (evaluated in the previous section) is related to $g_{N}$ (the coupling
constant of the bare state $N$ to $\overline{n}n$ in the case of no mixing)
by the relation 
\begin{equation}
g_{G^{\prime }}^{\overline{n}n}=M^{G^{\prime },N}g_{N}(p^{2}=M_{G^{\prime
}}^{2}),  \label{trenta}
\end{equation}
where $g_{N}(p^{2}=M_{G^{\prime }}^{2})$ is determined from (\ref{compcond})
evaluated at the physical mass with $p^{2}=M_{G^{\prime }}^{2}$. Eq. (\ref
{trenta}) states that the $\overline{n}n$ to $G^{\prime }$ coupling equals
the $\overline{n}n$ admixture in $G^{\prime }$ (which is the matrix element $%
M^{G^{\prime },N}$ we want to determine) times the $N$ to $\overline{n}n$
coupling evaluated on mass-shell of $G^{\prime }$.

Using (\ref{compcond}) and (\ref{pippo}) we obtain for $M^{G^{\prime },N}:$%
\begin{equation}
M^{G^{\prime },N}=\frac{g_{G^{\prime }}^{\overline{n}n}}{g_{N}(p^{2}=M_{G^{%
\prime }}^{2})}=sign(\alpha _{G^{\prime }}^{\overline{n}n})\left[ \left( 
\frac{\partial (T^{\overline{n}n,\overline{n}n})^{-1}}{\partial p^{2}}%
\right) \left( \frac{\partial \Sigma _{N}(p^{2})}{\partial p^{2}}\right)
\right] _{p^{2}=M_{G^{\prime }}^{2}}^{-1/2}.
\end{equation}
Generalizing this result for a generic component of the mixing matrix $M$ we
have 
\begin{equation}
M^{i,a}=\frac{g_{i}^{a}}{g_{a}(p^{2}=M_{i}^{2})}=sign(\alpha _{i}^{a})\left[
\left( \frac{\partial (T^{a,a})^{-1}}{\partial p^{2}}\right) \left( \frac{%
\partial \Sigma _{a}(p^{2})}{\partial p^{2}}\right) \right]
_{p^{2}=M_{i}^{2}}^{-1/2}.  \label{mixingmatrix}
\end{equation}
with $i=N^{\prime },G^{\prime },S^{\prime }$ and $a=N\equiv \overline{n}%
n,G\equiv gg,S\equiv \overline{s}s.$ Last expression, while not being based
on a rigorous derivation, can be regarded as a $definition$ of the mixing
matrix in a Bethe-Salpeter approach. We explicitly show in Appendix A for
the reduced problem of two mixed fields that this definition of the mixing
matrix is the analogous one the Klein-Gordon case.

Furthermore, the components of the mixed state $\left| i\right\rangle $ are
correctly normalized with 
\begin{equation}
1=(M^{iN})^{2}+(M^{iG})^{2}+(M^{iS})^{2}  \label{normmixmat}
\end{equation}
as verified both numerically and analytically (Appendix A). This is a
further confirmation of the consistency of the definition (\ref{mixingmatrix}%
). While the rows of $M$ are properly normalized, this does not hold for the
respective columns of $M$ because of the $p^{2}$ dependence of the $T$
matrix and of the mass operators. The rows are evaluated at different
on-shell values of $p^{2}$ values with $p^{2}=M_{i}^{2}.$ This implies that
the matrix $M$ is not orthogonal, but here we will demonstrate that
deviations from orthogonality are small for both choices of the quark
propagator.

\subsection{Comparison with other mixing schemes}

We again consider the Klein-Gordon case and its mass equation $%
Det[p^{2}+T_{K-G}]=0$ which reads: 
\begin{equation}
(p^{2}-M_{N}^{2})(p^{2}-M_{G}^{2})(p^{2}-M_{S}^{2})-f^{2}\left(
(p^{2}-M_{N}^{2})+2r^{2}(p^{2}-M_{S}^{2})\right) =0.  \label{kgmix}
\end{equation}

In our approach the mass equation $Det[T^{-1}]=0$ (see Eq. (\ref{t})) is
written out as: 
\[
(1-K_{N}\Sigma _{N}(p^{2}))(1-K_{G}\Sigma _{G}(p^{2}))(1-K_{S}\Sigma
_{S}(p^{2}))- 
\]
\[
-K_{SG}^{2}\{\Sigma _{G}(p^{2})\Sigma _{S}(p^{2})(1-K_{N}\Sigma _{N}(p^{2})) 
\]
\begin{equation}
+2\Sigma _{G}(p^{2})\Sigma _{S}(p^{2})(1-K_{S}\Sigma _{S}(p^{2}))\}=0
\label{bsextended}
\end{equation}

In order to compare the last expression with (\ref{kgmix}) we introduce the
functions $\eta _{a}(p^{2})$ as 
\begin{equation}
(1-K_{a}\Sigma _{a}(p^{2}))=(p^{2}-M_{a}^{2})\eta _{a}(p^{2})  \label{etas}
\end{equation}
for each $a=N,G,S$. The $\eta _{a}(p^{2})$ do not contain poles for $%
p^{2}=M_{a}^{2}$ since $(1-K_{a}\Sigma _{a}(p^{2}=M_{a}^{2}))=0$ (see (\ref
{kg}) and (\ref{mnms})). When substituting (\ref{etas}) in (\ref{bsextended}%
) we get: 
\begin{eqnarray}
&&(p^{2}-M_{N}^{2})(p^{2}-M_{G}^{2})(p^{2}-M_{S}^{2})-  \nonumber \\
- &&K_{SG}^{2}\frac{\Sigma _{G}(p^{2})\Sigma _{S}(p^{2})}{\eta
_{G}(p^{2})\eta _{S}(p^{2})}\left( (p^{2}-M_{N}^{2})+2\frac{\Sigma
_{N}(p^{2})}{\eta _{N}(p^{2})}\frac{\eta _{S}(p^{2})}{\Sigma _{S}(p^{2})}%
(p^{2}-M_{S}^{2})\right) =0  \label{etaeq}
\end{eqnarray}

Comparing (\ref{etaeq}) with (\ref{kgmix}) we deduce that the mixing
parameters $f$ and $r$ of the Klein-Gordon approach become $p^{2}$-dependent
functions in the case of composite scalar fields. In particular we make the
following identification 
\begin{eqnarray}
f^{2} &\rightarrow &f^{2}(p^{2})=K_{SG}^{2}\frac{\Sigma _{G}(p^{2})\Sigma
_{S}(p^{2})}{\eta _{G}(p^{2})\eta _{S}(p^{2})},  \nonumber \\
r^{2} &\rightarrow &r^{2}(p^{2})=\frac{\Sigma _{N}(p^{2})}{\eta _{N}(p^{2})}%
\frac{\eta _{S}(p^{2})}{\Sigma _{S}(p^{2})}.  \label{compkg}
\end{eqnarray}
It should be noted that the composite approach generates a value of $r\neq 1$
which reflects the deviation from the flavor blindness hypothesis. Although
we set up the quark-gluon interaction assuming this hypothesis, on the
composite hadronic level we obtain a breaking of this symmetry due to the
flavor dependence of the quark propagators. As we will see in our numerical
evaluation (see section IV and Figs. 5 and 6) the functions $f(p^{2})$ and $%
r(p^{2})$ vary slowly in the momentum range of interest, thus justifying $a$ 
$posteriori$ the Klein-Gordon approach.

Replacing in (\ref{tkg}) the constants $f$ and $r$ by the running functions $%
f(p^{2})$ and $r(p^{2})$ we have 
\begin{equation}
T_{K-G}\rightarrow T_{K-G}(p^{2})=\left( 
\begin{array}{lll}
-M_{N}^{2} & \sqrt{2}f(p^{2})r(p^{2}) & 0 \\ 
\sqrt{2}f(p^{2})r(p^{2}) & -M_{G}^{2} & f(p^{2}) \\ 
0 & f(p^{2}) & -M_{S}^{2}
\end{array}
\right) ,
\end{equation}
where the Klein-Gordon mass equation $Det[p^{2}+T_{K-G}(p^{2})]=0$ coincides
with the pole equation $Det[T^{-1}]=0$ by construction. When diagonalizing $%
T_{K-G}(p^{2})$ we obtain a momentum-dependent transition matrix $%
M_{K-G}(p^{2})$ $\subset SO(3)$ (for each $p^{2})$, which cannot be directly
interpreted as the mixing matrix. To extract the composition of a mixed
state one has to consider its on-shell mass value. For example, the mixing
coefficients for the diagonal state $\left| G^{\prime }\right\rangle $ are
obtained from the second row of $M_{K-G}(p^{2}=M_{G^{\prime }}^{2})$ at the
corresponding on-shell value. We then can define a mixing matrix $M^{\prime }
$ (the prime serves to distinguish it from $M$ defined in the previous
subsection) where we have $M^{\prime G^{\prime },a}=M_{K-G}^{G^{\prime
},a}(p^{2}=M_{G^{\prime }}^{2})$ $(a=N,G,S)$ and in general we get: 
\begin{equation}
M^{\prime i,a}=M_{K-G}^{i,a}(p^{2}=M_{i}^{2}).  \label{mixingmatrix2}
\end{equation}
Above procedure arises from the analogy to the Klein-Gordon case and
indicates that $M^{\prime }$ is a natural generalization of $M_{K-G}$,
provided that we evaluate the rotated states at their corresponding on-shell
mass value. Again, due to the $p^{2}$-dependence and the on-shell evaluation
the mixing matrix $M^{\prime }$ is not orthogonal, where in the limit of $%
f(p^{2})=const$ and $r(p^{2})=const$ one has $M^{\prime }=M_{K-G}$ as
desired. The numerical results of $M$ (\ref{mixingmatrix}) and $M^{\prime }$
(\ref{mixingmatrix2}) are very similar, as we will show in the next section.
The analytical argument leading to $M\sim M^{\prime }$ are presented in
Appendix A. Because of the weak $p^{2}$ dependence of the mixing functions $%
f(p^{2})$ and $r(p^{2}),$ the mixing matrix $M$ (and $M^{\prime }\sim M$)
will be almost orthogonal, as shown in section IV.

In the three-state mixing scheme many phenomenological approaches \cite
{close95,gutsche,close} refer to an interaction matrix, which is
diagonalized linearly with respect to the masses: 
\begin{equation}
T_{QM}=\left( 
\begin{array}{lll}
-M_{N} & \sqrt{2}f^{*}r^{*} & 0 \\ 
\sqrt{2}f^{*}r^{*} & -M_{G} & f^{*} \\ 
0 & f^{*} & -M_{S}
\end{array}
\right) .
\end{equation}
Here we introduce the parameters $f^{*}$ and $r^{*}$ to distinguish them
from those in (\ref{kgmix}). The resulting equation for the masses of the
mixed states is then: 
\begin{equation}
(\sqrt{p^{2}}-M_{N})(\sqrt{p^{2}}-M_{G})(\sqrt{p^{2}}-M_{S})-f^{*2}\left( (%
\sqrt{p^{2}}-M_{N})+2r^{*2}(\sqrt{p^{2}}-M_{S})\right) =0.  \label{lin1}
\end{equation}
To relate our approach to the linear mass case with parameters $f^{*}$ and $%
r^{*}$ we start from (\ref{etaeq}) and decompose $(p^{2}-M_{i}^{2})=(\sqrt{%
p^{2}}-M_{i})(\sqrt{p^{2}}+M_{i})$. Then we have 
\begin{eqnarray}
&&(\sqrt{p^{2}}-M_{N})(\sqrt{p^{2}}-M_{G})(\sqrt{p^{2}}-M_{S})-\frac{%
f^{2}(p^{2})}{(\sqrt{p^{2}}+M_{S})(\sqrt{p^{2}}+M_{G})}*  \nonumber \\
&&\left( (\sqrt{p^{2}}-M_{N})+2r^{2}(p^{2})\frac{(\sqrt{p^{2}}+M_{S})}{(%
\sqrt{p^{2}}+M_{G})}(\sqrt{p^{2}}-M_{S})\right) =0.  \label{lin2}
\end{eqnarray}
Comparing (\ref{lin2}) to (\ref{lin1}) we find the ''running'' behavior for $%
f^{*}$ and $r^{*}$ with 
\begin{eqnarray}
f^{*2} &\rightarrow &f^{*2}(p^{2})=\frac{f^{2}(p^{2})}{(\sqrt{p^{2}}+M_{S})(%
\sqrt{p^{2}}+M_{G})}  \nonumber \\
r^{*2} &\rightarrow &r^{*2}(p^{2})=r^{2}(p^{2})\frac{(\sqrt{p^{2}}+M_{S})}{(%
\sqrt{p^{2}}+M_{G})},  \label{complin}
\end{eqnarray}
where extra-factors occur when relating the quadratic to the linear mass
case. The consistent limit of our approach is actually the Klein-Gordon
case, but Eq. (\ref{complin}) yields a direct comparison to the linear mass
case, where many phenomenological approaches are working in.

\subsection{Decay into two photons}

In this work we also analyze the two-photon decay rates of the scalar
isoscalar $f_0$ mesons considered. The two-photon decay width constitutes a
crucial test to analyze the charge content of the scalar mesons \cite
{close,Kleefeld:2001ds,DeWitt:2003rs}. The glueball component does not
couple directly to the two-photon state and leads to a suppression of the
decay width when present in the mixed state.

For the decay of the $N^{\prime }\equiv f_{0}(1370)$ meson by its $\overline{%
u}u$ constituents into two photons we have to consider the triangle diagram
of Fig. 3. Analogous diagrams occur for the $d$ and $s$ flavors in the loop.

When evaluating the two-photon decay we will consider the first choice, that
is the free form of the quark propagators, which does not imply a
modification of the QED Ward identity, thus considerably simplifying the
problem. Since we deal with a non-local theory as introduced by the vertex
functions, care has to be taken concerning local gauge invariance. The
triangle diagram amplitude of Fig. 3 
\begin{equation}
M_{triangle}^{\mu \nu }=-i\int \frac{d^{4}k}{(2\pi )^{4}}\cdot
Tr[S_{n}(p_{1})\gamma ^{\mu }S_{n}(p_{2})\gamma ^{\nu }S_{n}(p_{3})]%
\widetilde{\Phi }(q^{2}),
\end{equation}
where $k_{1}$ and $k_{2}$ are the photon momenta and $p_{1}=k+k_{1},$ $%
p_{2}=k,$ $p_{3}=k-k_{2}$, is not gauge invariant.

The trace reads 
\begin{eqnarray}
&&\!Tr[(p\!\!\!\slash_{1}+\mu _{n})\gamma ^{\mu }(p\!\!\!\slash_{2}+\mu
_{n})\gamma ^{\nu }(p\!\!\!\slash_{3}+\mu _{n})]  \nonumber \\
&=&4\mu _{n}\left( (k_{1}^{\nu }k_{2}^{\mu }-(k_{1}\cdot k_{2})g^{\mu \nu
})\right) +4\mu _{n}(4k^{\mu }k^{\nu }-g^{\mu \nu }(k^{2}-\mu _{n}^{2}))
\end{eqnarray}
where we have omitted the terms proportional to $k_{1}^{\mu }$ and $%
k_{2}^{\nu }$ because they do not contribute to the decay. The first part of
the trace is clearly gauge invariant, while the second one is not. Gauging
in addition the non-local interaction term (for the procedure see \cite
{lyubo,ivanov,terning}) gives rise to the two extra diagrams of Fig. 4,
where the photons can now couple directly to the non-local vertex. Each
separate diagram is not gauge invariant, but the physical amplitude, which
is the sum of the three diagrams, is gauge invariant \cite{lyubo}. We can
find the gauge invariant contribution of the triangle diagram by considering
the shift 
\begin{eqnarray}
\gamma ^{\mu } &\rightarrow &\gamma _{\perp ,k_{1}}^{\mu }=\gamma ^{\mu
}-k_{1}^{\mu }\frac{k\!\!\!\slash_{1}}{k_{1}^{2}}, \\
\gamma ^{\nu } &\rightarrow &\gamma _{\perp ,k_{2}}^{\nu }=\gamma ^{\nu
}-k_{2}^{\nu }\frac{k\!\!\!\slash_{2}}{k_{2}^{2}}.
\end{eqnarray}
The term 
\begin{equation}
M_{triangle,\bot }^{\mu \nu }=-i\cdot \int \frac{d^{4}k}{(2\pi )^{4}}\cdot
Tr[S_{n}(p_{1})\gamma _{\perp ,k_{1}}^{\mu }S_{n}(p_{2})\gamma _{\perp
,k_{2}}^{\nu }S_{n}(p_{3})]\widetilde{\Phi }(q^{2})  \label{triangleort}
\end{equation}
is then by construction gauge invariant. As shown in \cite{lyubo}, which we
refer to for a careful analysis of this issue, this can be done for each
diagram and the total amplitude is 
\begin{equation}
M_{total}^{\mu \nu }=M_{triangle}^{\mu \nu }+M_{bubble}^{\mu \nu
}+M_{tadpole}^{\mu \nu }=M_{triangle,\bot }^{\mu \nu }+M_{bubble,\bot }^{\mu
\nu }+M_{tadpole,\bot }^{\mu \nu }
\end{equation}
where the gauge breaking terms cancel.

As already discussed in \cite{lyubo} and also checked for our case
numerically, the gauge invariant part of the triangle diagram $%
M_{triangle,\bot }^{\mu \nu }$ represents the by far dominant contribution
to the two-photon decay width. For the case of a Gaussian vertex function
the bubble and tadpole diagrams of Fig. 4 are further suppressed in the
order of $10^{-5}.$

\medskip We finally give the analytic expression for $M_{triangle,\bot
}^{\mu \nu }$%
\begin{equation}
M_{triangle,\bot }^{\mu \nu }=(k_{1}^{\nu }k_{2}^{\mu }-(k_{1}\cdot
k_{2})g^{\mu \nu })\cdot I,
\end{equation}
where $I=I_{1}+I_{2}$ is calculated from (\ref{triangleort}) with 
\begin{equation}
I_{1}=-i\cdot 4\mu _{n}\int \frac{d^{4}k}{(2\pi )^{4}}\frac{1}{%
(p_{1}^{2}-\mu _{n}^{2})(p_{2}^{2}-\mu _{n}^{2})(p_{3}^{2}-\mu _{n}^{2})}%
\widetilde{\Phi }(q^{2}),
\end{equation}
\begin{equation}
I_{2}=i\cdot 4\mu _{n}\int \frac{d^{4}k}{(2\pi )^{4}}\frac{\left( \frac{%
2k^{2}}{(k_{1}\cdot k_{2})}-8\frac{(k\cdot k_{1})(k\cdot k_{2})}{(k_{1}\cdot
k_{2})^{2}}\right) }{(p_{1}^{2}-\mu _{n}^{2})(p_{2}^{2}-\mu
_{n}^{2})(p_{3}^{2}-\mu _{n}^{2})}\widetilde{\Phi }(q^{2}).
\end{equation}
The term $I_{1}$ also occurs in the neutral pion decay into two photons,
whereas the additional term $I_{2}$ turns out to have opposite sign to $%
I_{1} $ and tends to lower the decay rate.

With above results we finally obtain for the two-photon decay width of the
mixed scalar meson state like $N^{\prime }$: 
\begin{equation}
\Gamma _{N^{\prime }\rightarrow 2\gamma }=\frac{\pi }{4}\alpha
^{2}M_{N^{\prime }}^{3}\left[ 2\left( \frac{g_{N^{\prime }}^{\overline{n}n}}{%
\sqrt{2}}\right) N_{c}(\frac{4}{9}+\frac{1}{9})(I)+2g_{N^{\prime }}^{%
\overline{s}s}N_{c}(\frac{1}{9})(I^{(s)})\right] ^{2}
\end{equation}
where $N_{c}=3$ is the number of colors and $I^{(s)}$ is obtained from $I$
by replacing $\mu _{n}$ with $\mu _{s}$ in the quark propagator. Only the
coupling constants $g_{N^{\prime }}^{\overline{n}n}$ and $g_{N^{\prime }}^{%
\overline{s}s}$ of the mixed field $N^{\prime }$ contribute to the decay
rate, while $g_{N^{\prime }}^{gg}$ does not because gluons do not couple
directly to photons. The extra-factor 2 in front of the coupling constants
comes from the exchange diagram. Analogous expressions follow for the decay
rates of the other two resonances $S^{\prime }$ and $G^{\prime }$.

\section{\protect\medskip Numerical results and discussion}

First we proceed to constrain the parameters of the model. The mass scale of
the non-strange quarkonia in the scalar meson nonet is set by the physical
state $a_{0}(1450)$. We therefore assume that the mass of the unmixed $N$
meson state is close to this value. Using $M_{N}=1.377$ GeV as an input,
which is the value deduced in the phenomenological analysis of Ref. \cite
{close}, the mass of the bare $S$ meson state and the elementary quark
coupling constant $K_{N}$ are fixed from Eq.(\ref{mnms}) once the cut-off $%
\Lambda $ is specified. The bare scalar glueball mass is predicted from
lattice calculations in the range of $1611\pm 30\pm 160$ MeV \cite
{michael12003}. The analysis of \cite{close} prefers a value of $1.443$ GeV,
which we use as an additional input. The bare glueball mass fixes in turn
the gluonic coupling $K_{G}$ of Eq. (\ref{kg}).

We now consider separately the phenomenological consequences of the two
proposed choices for the quark propagators.

\subsubsection{ Case 1, free quark propagator}

\paragraph{Mass spectrum:}

As discussed in Section II we consider large effective quark masses in order
to avoid poles in the integration. For the quark mass we choose the
threshold value $\mu _{n}=0.86$ $GeV$ to prevent unphysical on-shell
production of a quark-antiquark pair.

For the vertex function we choose a cut-off of $\Lambda =1.5$ $GeV$
comparable to the values chosen in the analysis of \cite{lyubo} for the
light meson sector. We will subsequently vary $\Lambda $ within reasonable
range, in order to check the dependence of the results on the specific
value. Actually, we find a remarkable stability of the results under changes
of $\Lambda $ as discussed further on. The mixing coefficient $K_{SG}$ and
the effective strange quark mass $\mu _{s}$ in the quark propagator are
fixed by the experimental masses $M_{G^{\prime }}=1507\pm 5$ $MeV$ and $%
M_{S^{\prime }}=1713\pm 7$ $MeV$. We then can compare $K_{SG}$ with
phenomenological approaches and with estimates from the lattice following
the expressions of (\ref{compkg}) and (\ref{complin}). Here we do not use
the mass of the resonance $N^{\prime }\equiv f_{0}(1370)$ as a further
constraint, since it is rather broad and ill-determined. For the optimal
choice of $\mu _{s}=0.989$ $GeV$ and $K_{SG}$ $=0.55$ $GeV^{-1}$ we obtain
for the masses of the physical $N^{\prime }$ and the bare S states: 
\begin{equation}
M_{S}=1.696~GeV,~~M_{N^{\prime }}=1.297~GeV.  \label{massresults}
\end{equation}
The obtained quark ''mass difference'' $\mu _{s}-\mu _{n}=129$ $MeV$ is
close to the upper limit of the current quark mass difference of about $130$
MeV. With these quark mass values the bare mass level scheme $%
M_{N}<M_{G}<M_{S}$ comes out naturally. A reversal of the bare scheme with $%
M_{N}<M_{S}<M_{G}$ would require a rather small mass difference $\mu
_{s}-\mu _{n}$ in conflict with phenomenology.

\paragraph{Mixing matrix:}

\medskip The mixing matrix $M$ linking the rotated and the unrotated states
is: 
\begin{equation}
\left( 
\begin{array}{l}
\left| N^{\prime }\right\rangle \\ 
\left| G^{\prime }\right\rangle \\ 
\left| S^{\prime }\right\rangle
\end{array}
\right) =\left( 
\begin{array}{lll}
0.80 & 0.59 & 0.10 \\ 
-0.60 & 0.76 & 0.26 \\ 
0.08 & -0.27 & 0.96
\end{array}
\right) \left( 
\begin{array}{l}
\left| N\right\rangle \\ 
\left| G\right\rangle \\ 
\left| S\right\rangle
\end{array}
\right)  \label{mixmat}
\end{equation}
showing a similar pattern as in the phenomenological work of \cite{close}:
the center state $\left| G^{\prime }\right\rangle $, identified with the $%
f_{0}(1500)$, has a dominant glueball component and the quark components $%
\left| N\right\rangle $ and $\left| S\right\rangle $ which are out of phase.
Latter effect causes destructive interference for the $K\bar{K}$ decay mode
consistent with experimental observation \cite{close95}. On the other hand $%
\left| N^{\prime }\right\rangle $ and $\left| S^{\prime }\right\rangle $
show a tendency to be dominated by the $\bar{n}n$ and $\bar{s}s$ constituent
components as also deduced in \cite{gutsche,close}.

As already indicated, the mixing matrix of (\ref{mixmat}) is not orthogonal.
The deviation from orthogonality is displayed by $M\cdot M^{t}$ (which is
just the identity in the Klein-Gordon limit): 
\begin{equation}
M\cdot M^{t}=\left( 
\begin{array}{lll}
1 & -0.0028 & -0.0019 \\ 
-0.0028 & 1 & -0.0074 \\ 
-0.0019 & -0.0074 & 1
\end{array}
\right) ,
\end{equation}
where the off-diagonal elements turn out to be very small. This in turn
implies that the Klein-Gordon limit is fully appropriate to set up the
mixing scheme of the scalar meson states. Furthermore, the unities on the
diagonal are in accord with (\ref{normmixmat}).

In Section III.C we have introduced the mixing matrix $M^{\prime }$. The
numerical evaluation yields 
\begin{equation}
M^{\prime }=\left( 
\begin{array}{lll}
0.79 & 0.60 & 0.09 \\ 
-0.61 & 0.75 & 0.24 \\ 
0.08 & -0.26 & 0.96
\end{array}
\right) ,
\end{equation}
where the result is nearly identical to the one for M. The reasons for this
coincidence are given in Appendix A.

The pattern observed in the mixing matrix only weakly depends on the
particular choice of $\Lambda $. Similarly, a change in the value of $\mu
_{n}$ up to $1.1$ $GeV$ alters the mixing pattern only within 5 \%. The weak
dependence of the results on $\Lambda $ and $\mu _{n}$ is explicitly
indicated in Appendix B.

\paragraph{Coupling constants:}

In the quantum mechanical approach the amplitude for the decay of the mixed
state $\left| i\right\rangle $ (with $i=N^{\prime },G^{\prime },S^{\prime })$
into two pions is fed by the $|N>$ component in the strong coupling limit 
\cite{close95}. Hence the $2\pi $ decay amplitude is proportional to $%
M^{i,N} $ defined in Eq. (\ref{mixingmatrix}). In this limit we obtain for
the ratio of $2\pi $ decay amplitudes $\left| A_{N^{\prime }\rightarrow 2\pi
}/A_{G^{\prime }\rightarrow 2\pi }\right| $ $\alpha $ $\left| M^{N^{\prime
},N}/M^{G^{\prime },N}\right| =0.80/0.60=1.33$ (in the preferred solution of 
\cite{close} one has for this ratio $1.13$). In our full approach the decay
amplitude is related to the respective coupling constant, hence we obtain
for the ratio $\left| A_{N^{\prime }\rightarrow 2\pi }/A_{G^{\prime
}\rightarrow 2\pi }\right| $ $\alpha $ $\left| g_{N^{\prime }}^{\overline{n}%
n}/g_{G^{\prime }}^{\overline{n}n}\right| =1.64$, which shows a clear
enhancement. This might explain why the state $N^{\prime }\equiv f_{0}(1370)$
is considerably broader than the $G^{\prime }\equiv f_{0}(1500)$. The strong
deviations of mixing coefficients and coupling constants is a non-negligible
effect, which can be traced to the fact that the coupling constants $%
g_{a}(p^{2})=\left( \partial \Sigma _{a}(p^{2})/\partial p^{2}\right)
^{-1/2} $ with $a=N,G,S$ are momentum dependent. If $g_{a}(p^{2})=const,$ we
would recover the quantum mechanical limit of $M^{N^{\prime
},N}/M^{G^{\prime },N}=g_{N^{\prime }}^{\overline{n}n}/g_{G^{\prime }}^{%
\overline{n}n}$ (and similarly for all the other ratios). Again, the
relation $\left| g_{N^{\prime }}^{\overline{n}n}/g_{G^{\prime }}^{\overline{n%
}n}\right| >\left| M^{N^{\prime },N}/M^{G^{\prime },N}\right| $ is stable
when changing the parameters $\Lambda $ and $\mu _{n}$.

The complete set of resulting coupling constants is summarized as 
\begin{equation}
\left( 
\begin{array}{c}
\begin{array}{l}
g_{N^{\prime }}^{\overline{n}n}:g_{G^{\prime }}^{\overline{n}n}:g_{S^{\prime
}}^{\overline{n}n}=7.35:-4.48:0.40 \\ 
g_{N^{\prime }}^{\overline{s}s}:g_{G^{\prime }}^{\overline{s}s}:g_{S^{\prime
}}^{\overline{s}s}=1.20:2.69:8.10
\end{array}
\\ 
g_{N^{\prime }}^{gg}:g_{G^{\prime }}^{gg}:g_{S^{\prime
}}^{gg}=0.91:0.95:-0.23
\end{array}
\right) .
\end{equation}

These results indicated here are of course model dependent, but they point
to an interesting aspect in the comparison with other studies: the ratio of
coupling constants, relevant for the strong two-body decay modes, can vary
rather sensibly from the ratios of mixing matrix elements. This deviation is
a consequence of the bound-state nature of the scalar mesons in a covariant
framework.

It would of course be interesting to calculate the decays into two
pseudoscalar mesons by loop diagrams directly. But this would require a
consistent knowledge of the quark propagators and a careful study of the
pseudoscalar meson-quark vertex functions to treat both the scalar mesons
above $1$ $GeV$ and the light pseudoscalar mesons in a unified way.

\paragraph{Running mixing functions:}

As already explained in the section devoted to the comparison with other
mixing schemes, here we obtain a momentum dependent mixing strength. We
developed two formulations, $f(p^{2})$ of Eq. (\ref{compkg}) and $%
f^{*}(p^{2})$ of Eq. (\ref{complin}), in order to compare to the
Klein-Gordon case and the quantum mechanical linear mass limit. Our results
for $p^{2}$-dependence of the mixing strength is summarized in Fig. 5. The
running function $f^{*}(p^{2})$ varies in the range between $60$ and $64$ $%
MeV$ for the $p^{2}$ values of interest. Our result should be compared to
the value of $85\pm 10$ $MeV$ obtained in \cite{close}. Lattice calculations
in the quenched approximation obtain $f^{*}=43\pm 31MeV$ \cite{weing,weing2}
with a large uncertainty but of similar magnitude. Other quantum mechanical
studies use fitted mixing parameters of $f^{*}=77MeV$ \cite{weing}, $%
f^{*}=64\pm 13MeV$ \cite{weing2} and $f^{*}=80$ $MeV$ \cite{gutsche}. Since $%
f^{*}(p^{2})$ does not vary drastically in the region considered, this weak
dependence explains why the mixing matrix $M$ (and analogously $M^{\prime })$
is ''almost'' orthogonal. In the effective QCD approach of Ref. \cite{simonov} 
a smaller quarkonia-glueball mixing is found; however, strong
mixing is obtained when including an intermediate hybrid state appearing
in the same mass region as the glueball and quarkonia ones.

Another characteristics of the non-local covariant approach is the dynamical
generation of flavour blindness breaking with $r>1$. The values of the
matched ''running'' functions $r(p^{2})$ and $r^{*}(p^{2})$ (Fig. 6) are by
10-20 percent larger than unity. The lattice result of $r^{*}=1.20\pm 0.07$ 
\cite{weing} is in rather good agreement with our evaluation.

The characteristics of the mixing parameters are again rather stable when
changing the parameters.

\paragraph{Two-photon decay widths:}

We first consider the decay widths of the bare states $N$ and $S$ when
mixing is neglected. In this case the coupling constants $N-\overline{n}n$
and $S-\overline{s}s$ are given by (\ref{compcond}). We obtain 
\begin{equation}
\Gamma _{N\rightarrow 2\gamma }=0.821~keV,~\Gamma _{S\rightarrow 2\gamma
}=0.08~~keV.
\end{equation}
As explained previously, the amplitude of the triangle diagram which gives
the dominant contribution to the decay is proportional to $(I_{1}+I_{2})$.
The extra term $I_{2}$, absent in the neutral pion case, has the opposite
sign to $I_{1}$ and the ratio $\left| I_{2}/I_{1}\right| $ grows with
increasing mass of the resonance. In the decay of the bare state $N$ the
term $I_{2}$ lowers the decay rate by a factor of $5.22.$

For the two-photon decays of the physical states, where the coupling
constants of the mixed states are used, we get: 
\begin{equation}
\Gamma _{N^{\prime }\rightarrow 2\gamma }=0.453~keV,~\Gamma _{G^{\prime
}\rightarrow 2\gamma }=0.273~KeV,~\Gamma _{S^{\prime }\rightarrow 2\gamma
}=0.125~keV,
\end{equation}
resulting in the ratios of 
\begin{equation}
\Gamma _{N^{\prime }\rightarrow 2\gamma }/\Gamma _{S^{\prime }\rightarrow
2\gamma }=3.62,~\Gamma _{G^{\prime }\rightarrow 2\gamma }/\Gamma _{S^{\prime
}\rightarrow 2\gamma }=2.19.
\end{equation}
which are similar to the ones obtained in \cite{close} with $\Gamma
_{N^{\prime }\rightarrow 2\gamma }/\Gamma _{S^{\prime }\rightarrow 2\gamma
}=3.56$ and $\Gamma _{G^{\prime }\rightarrow 2\gamma }/\Gamma _{S^{\prime
}\rightarrow 2\gamma }=2.36$. Current experimental upper limits for $%
G^{\prime }\equiv f_{0}(1500)$ and for $S^{\prime }\equiv f_{0}(1710)$ are 
\cite{pdg2002}: 
\begin{eqnarray}
\Gamma _{f_{0}(1500)\rightarrow 2\gamma }\frac{\Gamma
_{f_{0}(1500)\rightarrow \pi \pi }}{\Gamma _{f_{0}(1500)tot}} &<&0.46~keV, \\
\Gamma _{f_{0}(1710)\rightarrow 2\gamma }\frac{~\Gamma
_{f_{0}(1710)\rightarrow K\overline{K}}}{\Gamma _{f_{0}(1710)tot}} &<&0.110%
\text{ }keV~.
\end{eqnarray}

\medskip Multiplying our theoretical results by the experimental ratios \cite
{pdg2002} $\frac{\Gamma _{f_{0}(1500)\rightarrow \pi \pi }}{\Gamma
_{f_{0}(1500)tot}}=0.454\pm 0.104$ and $\frac{~\Gamma
_{f_{0}(1710)\rightarrow K\overline{K}}}{\Gamma _{f_{0}(1710)tot}}%
=0.38_{-0.19}^{+0.09}$ we find:

\begin{eqnarray}
(\Gamma _{f_{0}(1500)\rightarrow 2\gamma })_{theory}\frac{\Gamma
_{f_{0}(1500)\rightarrow \pi \pi }}{\Gamma _{f_{0}(1500)tot}} &=&0.124\pm
0.028\text{ }keV \\
(\Gamma _{f_{0}(1710)\rightarrow 2\gamma })_{theory}\frac{\Gamma
_{f_{0}(1710)\rightarrow K\overline{K}}}{\Gamma _{f_{0}(1710)tot}}
&=&0.0475_{-0.0237}^{+0.0112}\text{ }keV\text{,}
\end{eqnarray}
in accord with the experimental upper limits.

The experimental two-photon decay width of the scalar resonance $f_{0}(1370)$
has been seen; originally on \cite{pdg2000} two values were indicated, i.e. $%
3.8\pm 1.5$ $keV$ and $5.4\pm 2.3$ $keV.$ However, it is not clear if the
two-photon signal comes from the $f_{0}(1370)$ or from the high mass end of
the broad $f_{0}(400-1200).$ The PDG currently \cite{pdg2002,pdg2004} seems
to favor this last possibility, but it states in a footnote that this data
could also be valid for the $f_{0}(1370).$ We therefore interpret the two
experimental values as an upper limit for the two-photon decay width of the $%
f_{0}(1370).$ The result for $\Gamma _{N^{\prime }\equiv
f_{0}(1370)\rightarrow 2\gamma }$ is an order of magnitude smaller than
these upper limits. A precise experimental determination of the two-photon
decay values of these scalar states would clearly help in understanding
their structure.

For what concerns the cut-off dependence of the decay rates, we note that an
increase in the cut-off leads to a weak increase of the decay widths as
indicated in Appendix B, but the ratios remain stable.

\subsubsection{Case 2, entire function}

In the following we summarize our results for the quark propagator of Eq. (%
\ref{propent}), described by an entire function modelling confinement.

\paragraph{Mass spectrum:}

We fix the parameters in the same fashion as for the previous case (that is
to $M_{N}=1.377$ $GeV,$ $M_{G}=1.443$ $GeV,$ $M_{G^{\prime }}=1.507$ $GeV$%
and $M_{S^{\prime }}=1.713$ $GeV$), but this time we have the parameter $%
\beta $ of the quark propagator of Eq. (\ref{propent}) together with the
mixing strength $K_{SG}$. For an identical cut-off with $\Lambda =1.5$ $GeV$
we get: 
\begin{equation}
M_{S}=1.698~GeV,~M_{N^{\prime }}=1.297~GeV  \label{massent}
\end{equation}
obtained for the fit values of $\beta =2.44$ $GeV^{-2}$ and $K_{SG}$ $=0.284$
$GeV^{-1}.$ The two results of Eq. (\ref{massent}) are essentially identical
to the previous case of Eq. (\ref{massresults}). Also in this case the
reversed bare level ordering with $M_{N}<M_{S}<M_{G}$ is disfavored,
requiring a decrease of $\beta $ of the order of $10^{-2}$. This is in
contrast to the requirement that the propagator behaves as a free in the
limit of small Minkowsky or Euclidean momenta.

\paragraph{Mixing matrix:}

The mixing matrix $M$ linking the rotated and the unrotated states is
practically unchanged when compared to case 1: 
\begin{equation}
\left( 
\begin{array}{l}
\left| N^{\prime }\right\rangle \\ 
\left| G^{\prime }\right\rangle \\ 
\left| S^{\prime }\right\rangle
\end{array}
\right) =\left( 
\begin{array}{lll}
0.80 & 0.59 & 0.10 \\ 
-0.59 & 0.76 & 0.25 \\ 
0.07 & -0.27 & 0.96
\end{array}
\right) \left( 
\begin{array}{l}
\left| N\right\rangle \\ 
\left| G\right\rangle \\ 
\left| S\right\rangle
\end{array}
\right) ,
\end{equation}
where $M\cdot M^{t}$ is very close to the identity matrix.

\paragraph{Coupling constants:}

The results concerning the admixture and coupling constant ratios is
analogous to the previous case. For the ratio of mixing amplitudes we get $%
\left| M^{N^{\prime },N}/M^{G^{\prime },N}\right| =0.79/0.59=1.36$, while
for the coupling constants we have $\left| g_{N^{\prime }}^{\overline{n}%
n}/g_{G^{\prime }}^{\overline{n}n}\right| =1.65$, again almost unchanged
with respect to case 1. The complete table reads: 
\begin{equation}
\left( 
\begin{array}{c}
\begin{array}{l}
g_{N^{\prime }}^{\overline{n}n}:g_{G^{\prime }}^{\overline{n}n}:g_{S^{\prime
}}^{\overline{n}n}=3.82:-2.32:0.22 \\ 
g_{N^{\prime }}^{\overline{s}s}:g_{G^{\prime }}^{\overline{s}s}:g_{S^{\prime
}}^{\overline{s}s}=0.63:1.37:4.11
\end{array}
\\ 
g_{N^{\prime }}^{gg}:g_{G^{\prime }}^{gg}:g_{S^{\prime
}}^{gg}=0.91:0.95:-0.23
\end{array}
\right) .
\end{equation}

\paragraph{Running mixing functions}

Results for $f(p^{2})$, $f^{*}(p^{2}),$ $r(p^{2})$ and $r^{*}(p^{2})$ are
summarized in Fig. 6, with the same quantitative behavior as in Fig. 5.

We conclude this section by noting that the quark propagator, modelling
confinement, gives rise to very similar results as the free propagator with
a large effective quark mass. Again, changes in the cutoff $\Lambda $ do not
alter the qualitative features of the results. This result is rather
encouraging, since the model predictions considered here, seemingly do not
depend on the particular choice of the quark propagator.

\section{\protect\medskip Conclusions}

\medskip In this work we utilized a covariant constituent approach to
analyze glueball-quarkonia mixing in the scalar meson sector above $1$ $GeV$%
. We used simple forms for the quark and gluon propagators in order to avoid
unphysical threshold production of quarks and gluons. Although quark and
gluon propagators are directly accessible in lattice simulations in the
Euclidean region, an extrapolation to the Minkowsky region, also needed
here, is not straightforward. We therefore considered relatively simple
choices of the propagators, which allowed us to point out some features of
the mixing of covariant bound states. We tried to work out similarities and
differences with phenomenological approaches, in particular with respect to
the analysis of \cite{close}. Although in a covariant approach the mixing
matrix is in general not orthogonal, in the present case only small
deviations from orthogonality are obtained. This in turn leads to a mixing
pattern rather similar to that of Ref. \cite{close}. The mixing matrix $M$
has been introduced by knowledge of the coupling constants of the mixed and
unmixed states taking into account their $p^{2}$ dependence. The resulting
matrix $M$ is analogous to the Klein-Gordon case as shown both numerically
and in part analytically.

Many properties we analyzed, such as the appearance of the ''running''
mixing parameters $f(p^{2})$ and $r(p^{2})$, are rather independent on the
choice of the particular quark propagator. The numerical results for $%
f(p^{2})$ and $r(p^{2})$ are in qualitative accord with the lattice
evaluations. We generate a dynamical breaking of the glueball flavor
blindness corresponding to $r$ slightly bigger than unity, which is directly
connected to the isospin symmetry violation at the level of the quark
propagators. Again, all these considerations do not depend on the choice of
the two proposed quark propagator forms and on the employed parameter sets.
Another interesting result of our approach is that the bare level ordering $%
M_{N}<M_{G}<M_{S}$ is naturally favored for both propagator choices.

As a further application we also evaluated the two-photon decay rates in the
context of the mixing model. The predicted results are in accord with the
present experimental upper limits. The respective ratios are also in
agreement with the phenomenological estimate of \cite{close}.

{\bf Acknowledgments}

\vspace*{.5cm}

\noindent We thank V. E. Lyubovitskij for intensive discussions on the
formalism. This work was supported by the Deutsche Forschungsgemeinschaft
(DFG) under contracts FA67/25-3 and GRK683.

\appendix

\section{Mixing matrices $M$ and $M^{\prime }$}

In Section III.C we introduced the mixing matrix $M$ (see eq. (\ref
{mixingmatrix})) and then in Section III.D alternatively $M^{\prime }$ (see
eq. (\ref{mixingmatrix2})). For both mixing matrices, as indicated in
Section IV, we find similar results. To demonstrate the close connection
between $M$ and $M^{\prime }$ we consider the reduced problem for the mixing
of two fields, $G$ and $S$, thus leaving out the field $N$ from the
discussion. In this case we have the rotated fields $G^{\prime }$ and $%
S^{\prime }$ only, and for $G^{\prime }$ we have $\left| G^{\prime
}\right\rangle =M^{G^{\prime }G}\left| G\right\rangle +M^{G^{\prime
}S}\left| S\right\rangle .$ $K$ and $\Sigma $ are now 2x2 matrices with: 
\begin{equation}
K=\left( 
\begin{array}{ll}
K_{G} & K_{SG} \\ 
K_{SG} & K_{S}
\end{array}
\right) ,\text{ }\Sigma =\left( 
\begin{array}{ll}
\Sigma _{G} & 0 \\ 
0 & \Sigma _{S}
\end{array}
\right) .
\end{equation}
Then we write the full expression for the matrix $T^{-1}$: 
\begin{equation}
T^{-1}=-(K^{-1}-\Sigma )=\left( 
\begin{array}{ll}
-\frac{K_{S}}{Det[K]}+\Sigma _{G} & \frac{K_{SG}}{Det[K]} \\ 
\frac{K_{SG}}{Det[K]} & -\frac{K_{G}}{Det[K]}+\Sigma _{S}
\end{array}
\right) ,
\end{equation}
from which we get 
\begin{equation}
T=\frac{1}{Det[T^{-1}]}\left( 
\begin{array}{ll}
-\frac{K_{G}}{Det[K]}+\Sigma _{S} & -\frac{K_{SG}}{Det[K]} \\ 
-\frac{K_{SG}}{Det[K]} & -\frac{K_{S}}{Det[K]}+\Sigma _{G}
\end{array}
\right) .  \label{treduced}
\end{equation}
The coupling constant $g_{G^{\prime }}^{gg}$ (\ref{pippo}) is then
explicitly given as 
\begin{eqnarray}
g_{G^{\prime }}^{gg} &=&\lim_{p^{2}\rightarrow M_{G^{\prime }}^{2}}\sqrt{%
(p^{2}-M_{G^{\prime }}^{2})T^{gg,gg}}=\lim_{p^{2}\rightarrow M_{G^{\prime
}}^{2}}\sqrt{\frac{(p^{2}-M_{G^{\prime }}^{2})}{Det[T^{-1}]}(-\frac{K_{G}}{%
Det[K]}+\Sigma _{S})}= \\
&=&\left( \sqrt{\left( \frac{\partial Det[T^{-1}]}{\partial p^{2}}\right)
^{-1}\left( -\frac{K_{G}}{Det[K]}+\Sigma _{S}\right) }\right)
_{p^{2}=M_{G^{\prime }}^{2}}.
\end{eqnarray}
The expressions for the other coupling constants follow from (\ref{pippo}).
Similarly, the explicit expression for $M^{G^{\prime }G}$ is 
\begin{eqnarray}
M^{G^{\prime }G} &=&\sqrt{g_{G^{\prime }}^{gg}\left( \frac{\partial \Sigma
_{G}}{\partial p^{2}}\right) _{p^{2}=M_{G^{\prime }}^{2}}}=  \nonumber \\
&=&\left( \sqrt{\left( \frac{\partial Det[T^{-1}]}{\partial p^{2}}\right)
^{-1}\left( -\frac{K_{G}}{Det[K]}+\Sigma _{S}\right) \left( \frac{\partial
\Sigma _{G}}{\partial p^{2}}\right) }\right) _{p^{2}=M_{G^{\prime }}^{2}}.
\label{mmreduced}
\end{eqnarray}
One has then similar relations for the other elements.

First we show that the components of the state $\left| G^{\prime
}\right\rangle =M^{G^{\prime }G}\left| G\right\rangle +M^{G^{\prime
}S}\left| S\right\rangle $ are correctly normalized. Using (\ref{mmreduced})
and (\ref{treduced}) we have 
\begin{eqnarray}
&&1=(M^{G^{\prime }G})^{2}+(M^{G^{\prime }S})^{2}=  \nonumber \\
&=&\left[ \left( \frac{\partial Det[T^{-1}]}{\partial p^{2}}\right)
^{-1}\left( \left( -\frac{K_{G}}{Det[K]}+\Sigma _{S}\right) \frac{\partial
\Sigma _{G}}{\partial p^{2}}+\left( -\frac{K_{S}}{Det[K]}+\Sigma _{G}\right) 
\frac{\partial \Sigma _{S}}{\partial p^{2}}\right) \right]
_{p^{2}=M_{G^{\prime }}^{2}}.
\end{eqnarray}
We come now to the equivalence of $M$ and $M^{\prime }.$ Let us consider the
ratio $\xi =\left| M^{G^{\prime }G}/M^{G^{\prime }S}\right| ,$ which can be
calculated explicitly from the basic definition (\ref{mixingmatrix}) and
form (\ref{treduced}) : 
\begin{eqnarray}
\xi &=&\left| \frac{M^{G^{\prime }G}}{M^{G^{\prime }S}}\right| =\left( \sqrt{%
\frac{-\frac{K_{G}}{Det[K]}+\Sigma _{S}}{-\frac{K_{S}}{Det[K]}+\Sigma _{G}}%
\left( \frac{\partial \Sigma _{G}/\partial p^{2}}{\partial \Sigma
_{S}/\partial p^{2}}\right) }\right) _{p^{2}=M_{G^{\prime }}^{2}}= \\
&=&\left( \left( \frac{-K_{G}+Det[K]\cdot \Sigma _{S}}{K_{SG}}\right) \sqrt{%
\frac{\partial \Sigma _{G}/\partial p^{2}}{\partial \Sigma _{S}/\partial
p^{2}}}\right) _{p^{2}=M_{G^{\prime }}^{2}},  \label{csi}
\end{eqnarray}
where the last term has been obtained making use of the equation $%
Det[T^{-1}]=0,$ which holds for $p^{2}=M_{G^{\prime }}^{2}.$

When we consider the analogous ratio calculated from the elements of $%
M^{\prime }$ (\ref{mixingmatrix}) (in this case we just have the running
parameter $f(p^{2})$ from equation (\ref{compkg}) and not $r(p^{2})$) we
find: 
\begin{equation}
\xi ^{\prime }=\left| \frac{M^{\prime G^{\prime }G}}{M^{\prime G^{\prime }S}}%
\right| =\frac{M_{S}^{2}-M_{G^{\prime }}^{2}}{K_{SG}}\sqrt{\left( \frac{\eta
_{G}(p^{2})}{\Sigma _{G}(p^{2})}\frac{\eta _{S}(p^{2})}{\Sigma _{S}(p^{2})}%
\right) _{p^{2}=M_{G^{\prime }}^{2}}}.
\end{equation}
We want to prove that $\xi \simeq \xi ^{\prime }$ (i.e. $M\simeq M^{\prime }$%
) in the weak mixing limit ($K_{SG}$ small) and in the $f(p^{2})=const$
limit, for which $M^{\prime }=M_{K-G.}$

A small mixing strength $K_{SG}$ implies that we can neglect the term
proportional to $K_{SG}^{2}$ in the determinant of equation (\ref{csi}).
Introducing the quantity $\eta _{S}(p^{2})$ of Eq. (\ref{etas}) we find: 
\begin{equation}
\xi \simeq \frac{K_{G}(M_{S}^{2}-M_{G^{\prime }}^{2})\eta _{S}(p^{2})}{K_{SG}%
}\sqrt{\left( \frac{\partial \Sigma _{G}(p^{2})/\partial p^{2}}{\partial
\Sigma _{S}(p^{2})/\partial p^{2}}\right) _{p^{2}=M_{G^{\prime }}^{2}}}.
\end{equation}
For $p^{2}\simeq M_{G}^{2}$ the following approximation is valid: 
\begin{equation}
\frac{\eta _{G}(p^{2})}{\Sigma _{G}(p^{2})}\simeq K_{G}^{2}\Sigma
_{G}^{\prime }(p^{2}).
\end{equation}
In fact, when $K_{SG}$ is small, $M_{G^{\prime }}^{2}$ is close to $%
M_{G}^{2} $. This allows us to write 
\begin{equation}
K_{G}\sqrt{\Sigma _{G}^{\prime }(p^{2}=M_{G^{\prime }}^{2})}\simeq \sqrt{%
\left( \frac{\eta _{G}(p^{2})}{\Sigma _{G}(p^{2})}\right)
_{p^{2}=M_{G^{\prime }}^{2}}}.  \label{csicond1}
\end{equation}

Let us now consider the second limit, for which $f(p^{2})=$ $const$ in the
region of interest, i.e. between $M_{G^{\prime }}^{2}$ and $M_{S^{\prime
}}^{2}$ in the 2 field case, and between $M_{N^{\prime }}^{2}$ and $%
M_{S^{\prime }}^{2}$ in the 3 field case. The condition $f(p^{2})=$ $const$
is satisfied if $\Sigma _{a}(p^{2})/\eta _{a}(p^{2})=c_{a},$ where $c_{a}$
is a constant for $a=G,S.$ In this case one has $M^{\prime }=M_{K-G},$ which
is orthogonal. We are then considering the orthogonal limit for $M^{\prime
}. $ The condition $\Sigma _{a}(p^{2})/\eta _{a}(p^{2})=c_{a}$ for the case $%
a=S $ implies the following form for $\Sigma _{S}(p^{2}):$%
\begin{equation}
\Sigma _{S}(p^{2})=\frac{c_{S}}{(p^{2}-M_{S}^{2})+c_{S}K_{S}}.
\label{sigmaapprox}
\end{equation}
This is of course an approximate form for $\Sigma _{S}(p^{2})$ valid in the
limit of a constant $c_{S}$. Note that the condition $\Sigma
_{S}(p^{2}=M_{S}^{2})-1/K_{S}=0$ from Eq. (2) is fulfilled. With this form
for $\Sigma _{S}(p^{2})$ we have 
\begin{equation}
\frac{\eta _{S}(p^{2})}{\sqrt{\partial \Sigma _{S}(p^{2})/\partial p^{2}}}=%
\sqrt{\frac{\eta _{S}(p^{2})}{\Sigma _{S}(p^{2})}}=\frac{1}{\sqrt{c_{S}}}
\label{csicond2}
\end{equation}
which is valid in the interval where (\ref{sigmaapprox}) is valid.

Plugging the approximations (\ref{csicond1}) and (\ref{csicond2}) in (\ref
{csi}) we have indeed 
\begin{equation}
\xi =\xi ^{\prime }=\frac{M_{S}^{2}-M_{G^{\prime }}^{2}}{K_{SG}}\sqrt{\left( 
\frac{\eta _{G}(p^{2})}{\Sigma _{G}(p^{2})}\frac{\eta _{S}(p^{2})}{\Sigma
_{S}(p^{2})}\right) _{p^{2}=M_{G^{\prime }}^{2}}}=\frac{M_{S}^{2}-M_{G^{%
\prime }}^{2}}{K_{SG}\sqrt{c_{S}c_{G}}},
\end{equation}
thus having similar results for $M$ and $M^{\prime }=M_{K-G}$ for the used
approximations. As shown in section IV, in the three field mixing case,
similar matrices are found. This is then true for all the parameters studied
in our work and for both propagator choices.

Note that the two conditions discussed in this appendix are satisfied: $%
K_{SG}$ is small (corresponding to a small difference $M_{S^{\prime
}}^{2}-M_{S}^{2}$ ) and the function $f(p^{2})$ is almost constant in the
momentum interval between $M_{N^{\prime }}^{2}$ and $M_{S^{\prime }}^{2}$
(see Fig. 5 and Fig. 6).

\section{Results for parameter variation in the case of the effective free
quark propagator}

To explicitly indicate the dependence on the cut-off value of the vertex
function here we summarize our results for $\Lambda =2~GeV$ with the fit
values $\mu _{s}=0.985$ $GeV$, $K_{SG}=0.122$ $GeV^{-1}$. The masses of
physical $N^{\prime }\equiv f_{0}(1370)$ and bare $S$ state are $%
M_{N^{\prime }}=1.287$ $GeV$ and $M_{S}=1.696$ $GeV$. 

Mixing matrix: 
\begin{equation}
M=\left( 
\begin{array}{lll}
0.79 & 0.61 & 0.11 \\ 
-0.61 & 0.74 & 0.27 \\ 
0.08 & -0.28 & 0.95
\end{array}
\right) .
\end{equation}
Set of coupling constants:

\begin{equation}
\left( 
\begin{array}{c}
\begin{array}{l}
g_{N^{\prime }}^{\overline{n}n}:g_{G^{\prime }}^{\overline{n}n}:g_{S^{\prime
}}^{\overline{n}n}=5.56:-3.65:0.35 \\ 
g_{N^{\prime }}^{\overline{s}s}:g_{G^{\prime }}^{\overline{s}s}:g_{S^{\prime
}}^{\overline{s}s}=0.94:2.04:6.24
\end{array}
\\ 
g_{N^{\prime }}^{gg}:g_{G^{\prime }}^{gg}:g_{S^{\prime
}}^{gg}=0.65:0.69:-0.20
\end{array}
\right) .
\end{equation}
\newline
Two-photon decay widths: 
\begin{eqnarray}
&& 
\begin{array}{l}
\Gamma _{N\rightarrow 2\gamma }=0.897keV
\end{array}
\text{ }\Gamma _{S\rightarrow 2\gamma }=0.095\text{ }keV \\
&& 
\begin{array}{lll}
\begin{array}{l}
\Gamma _{N^{\prime }\rightarrow 2\gamma }=0.465\text{ }keV
\end{array}
& 
\begin{array}{l}
\Gamma _{G^{\prime }\rightarrow 2\gamma }=0.326\text{ }keV
\end{array}
& 
\begin{array}{l}
\Gamma _{S^{\prime }\rightarrow 2\gamma }=0.152\text{ }keV
\end{array}
,
\end{array}
\\
&& 
\begin{array}{ll}
\Gamma _{N^{\prime }\rightarrow 2\gamma }/\Gamma _{S^{\prime }\rightarrow
2\gamma }=3.05 & \Gamma _{G^{\prime }\rightarrow 2\gamma }/\Gamma
_{S^{\prime }\rightarrow 2\gamma }=2.14
\end{array}
.
\end{eqnarray}

\medskip For an even further increase value of the cut-off width $\Lambda
=2.5$ $GeV$ $(\mu _{s}=0.982$ $GeV;$ $K_{SG}=0.041$ $GeV^{-1})$ we get $%
M_{N^{\prime }}=1.287$ $GeV$ and $M_{S}=1.695$ $GeV.$

Mixing matrix:

\begin{equation}
M=\left( 
\begin{array}{lll}
0.78 & 0.61 & 0.11 \\ 
-0.62 & 0.74 & 0.27 \\ 
0.08 & -0.29 & 0.95
\end{array}
\right) ,
\end{equation}
Set of coupling constants: 
\begin{equation}
\left( 
\begin{array}{c}
\begin{array}{l}
g_{N^{\prime }}^{\overline{n}n}:g_{G^{\prime }}^{\overline{n}n}:g_{S^{\prime
}}^{\overline{n}n}=4.64:-3.18:0.32 \\ 
g_{N^{\prime }}^{\overline{s}s}:g_{G^{\prime }}^{\overline{s}s}:g_{S^{\prime
}}^{\overline{s}s}=0.79:1.70:5.27
\end{array}
\\ 
g_{N^{\prime }}^{gg}:g_{G^{\prime }}^{gg}:g_{S^{\prime
}}^{gg}=0.51:0.55:-0.17
\end{array}
\right) .
\end{equation}
\newline
Two-photon decay widths: 
\begin{eqnarray}
&&
\begin{array}{l}
\Gamma _{N\rightarrow 2\gamma }=0.911\text{ }keV
\end{array}
,\Gamma _{S\rightarrow 2\gamma }=0.101\text{ }keV, \\
&&
\begin{array}{lll}
\begin{array}{l}
\Gamma _{N^{\prime }\rightarrow 2\gamma }=0.455\text{ }keV
\end{array}
& 
\begin{array}{l}
\Gamma _{G^{\prime }\rightarrow 2\gamma }=0.347\text{ }keV
\end{array}
& 
\begin{array}{l}
\Gamma _{S^{\prime }\rightarrow 2\gamma }=0.165\text{ }keV
\end{array}
,
\end{array}
\\
&&
\begin{array}{ll}
\Gamma _{N^{\prime }\rightarrow 2\gamma }/\Gamma _{S^{\prime }\rightarrow
2\gamma }=2.99 & \Gamma _{G^{\prime }\rightarrow 2\gamma }/\Gamma
_{S^{\prime }\rightarrow 2\gamma }=2.28.
\end{array}
\end{eqnarray}

As a last point we consider the increase of the effective quark mass
parameter $\mu _{n}$ from $0.86$ to $1.1$ $GeV$ in order to check its
influence on the results. For $\Lambda =1.5$ $GeV$ we get $\mu _{s}=1.2$ $%
GeV $ and $K_{SG}=1.00$ $GeV^{-1}.$ The masses are $M_{N^{\prime }}=1.307$ $%
GeV$ and $M_{S}=1.695$ $GeV$. The mixing matrix is

\begin{equation}
\text{ }M=\left( 
\begin{array}{lll}
0.81 & 0.57 & 0.09 \\ 
-0.57 & 0.78 & 0.25 \\ 
0.07 & -0.27 & 0.96
\end{array}
\right) ,
\end{equation}
where again no decisive variation from the previous cases is seen.


\medskip

\medskip

\medskip

\begin{figure}[htb]
\begin{center}
\resizebox{0.55\textwidth}{!}{\includegraphics{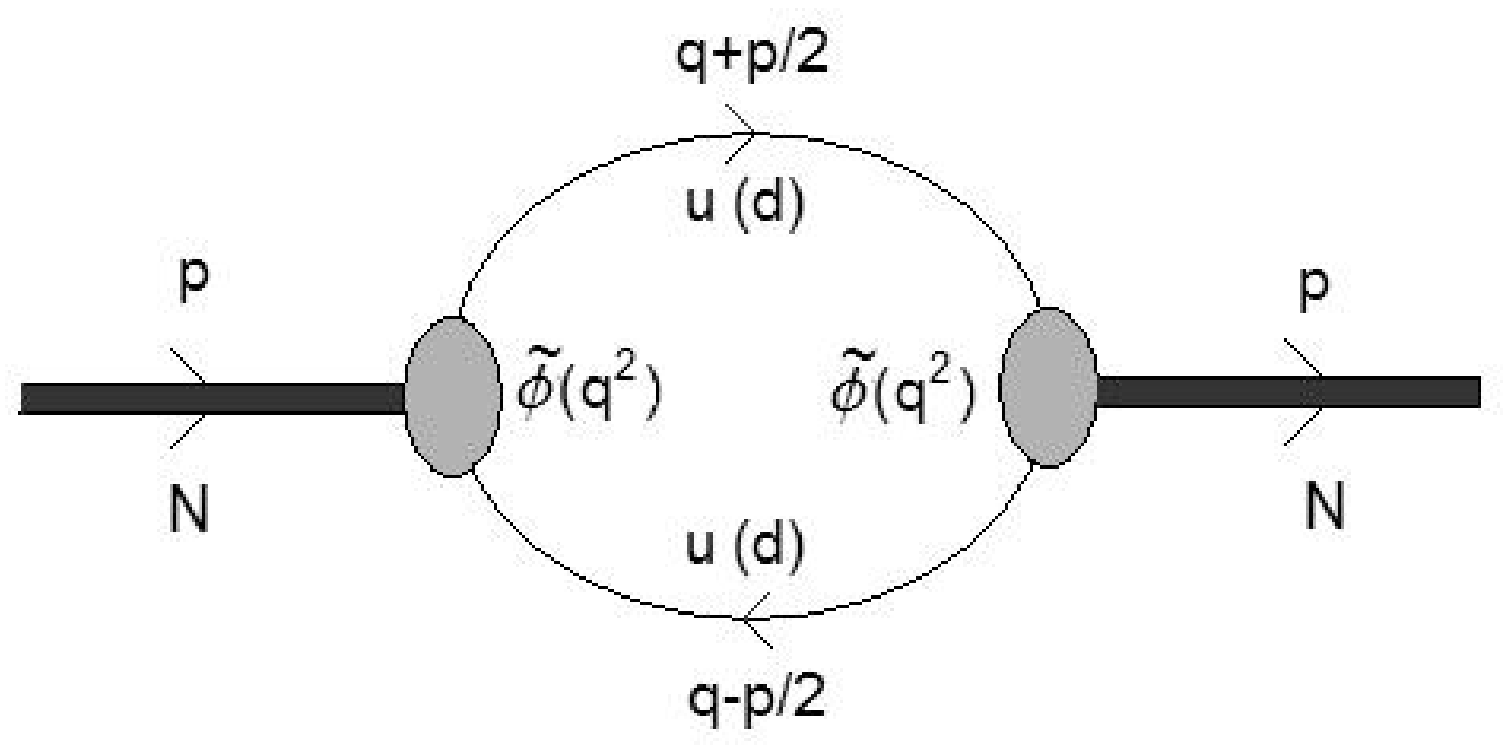}}
\caption{N-meson mass operator diagram. }
\label{fig1}
\end{center}
\end{figure}
\begin{figure}[htb]
\begin{center}
\resizebox{0.55\textwidth}{!}{\includegraphics{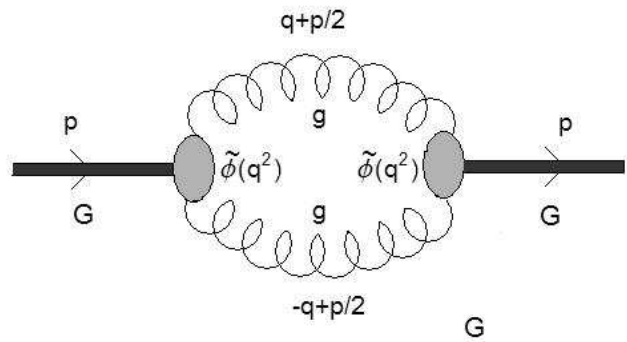}}
\caption{Glueball mass operator diagam. }
\label{fig2}
\end{center}
\end{figure}
\begin{figure}[htb]
\begin{center}
\resizebox{0.55\textwidth}{!}{\includegraphics{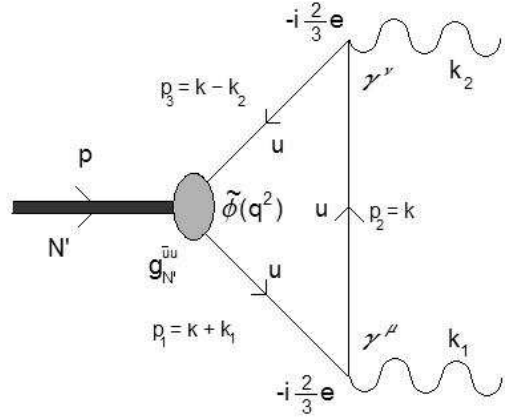}}
\caption{$N^{\prime }\equiv f_{0}(1370)$ meson decaying into two photons
through a u-loop.}
\label{fig3}
\end{center}
\end{figure}
\begin{figure}[htb]
\begin{center}
\resizebox{0.55\textwidth}{!}{\includegraphics{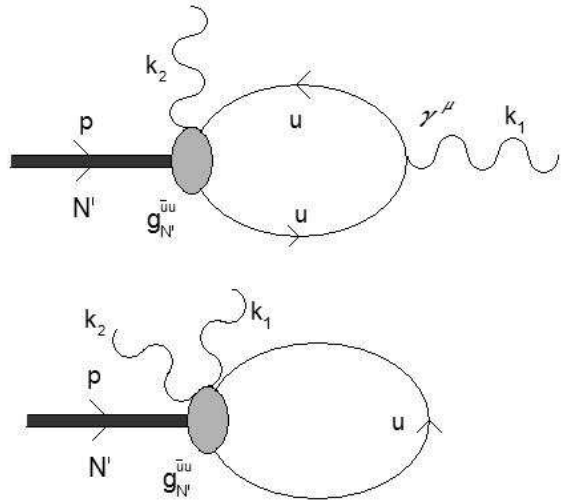}}
\caption{Bubble and tadpole diagrams for the two-photon decay of the
$N^{\prime }\equiv f_{0}(1370)$ meson. Both diagrams are suppressed,
but conceptually important for conservation of gauge invariance.}
\label{fig4}
\end{center}
\end{figure}
\begin{figure}[htb]
\begin{center}
\resizebox{0.75\textwidth}{!}{\includegraphics{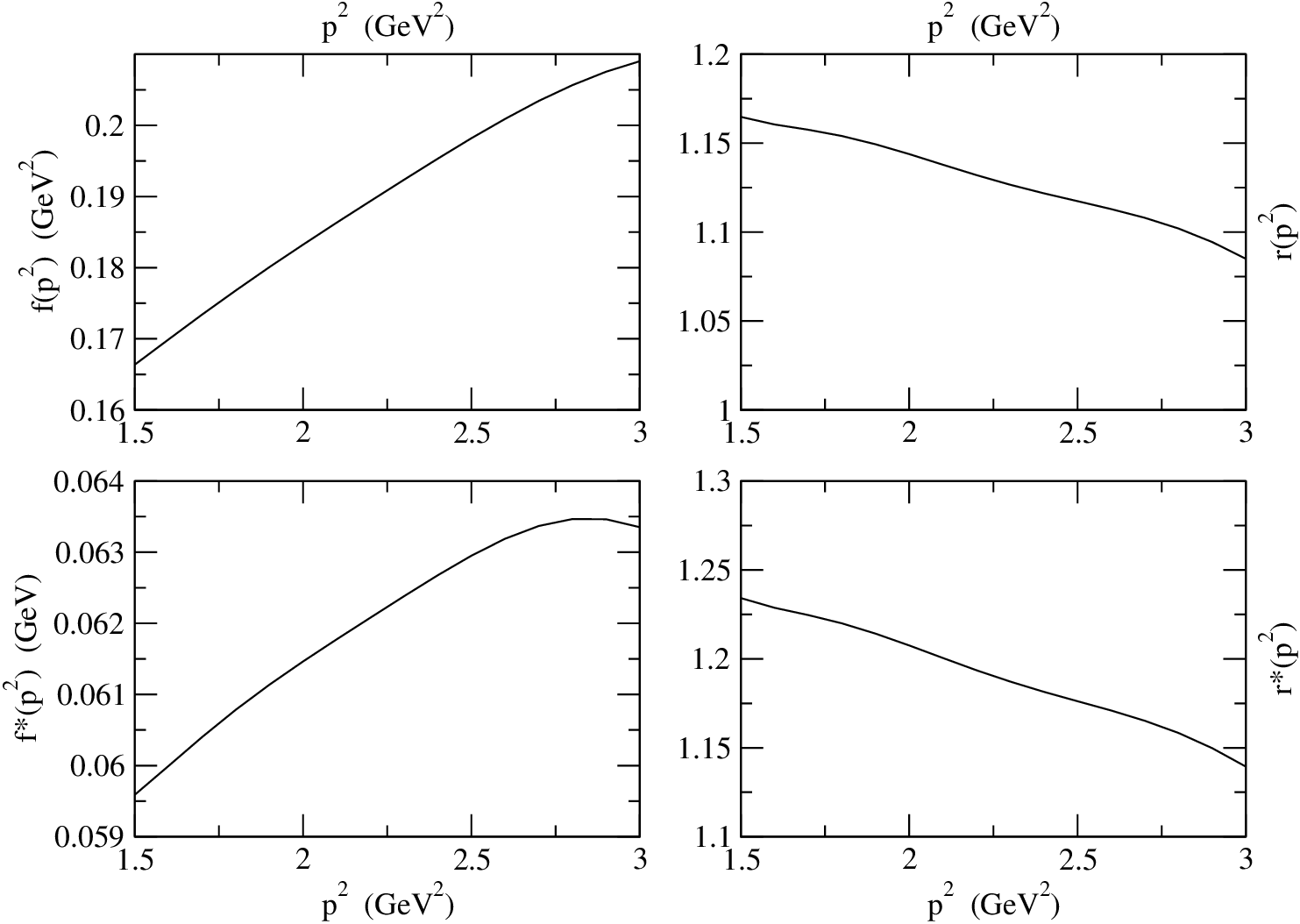}}
\caption{Plot of the mixing functions $f(p^{2}),r(p^{2})$ (eq. (\ref{compkg}%
)) and $f^{*}(p^{2}),r^{*}(p^{2})$ (eq. (\ref{complin})) for the
first propagator choice and with a vertex cut-off $\Lambda =1.5$ $GeV.$}
\label{fig5}
\end{center}
\end{figure}
\begin{figure}[htb]
\begin{center}
\resizebox{0.75\textwidth}{!}{\includegraphics{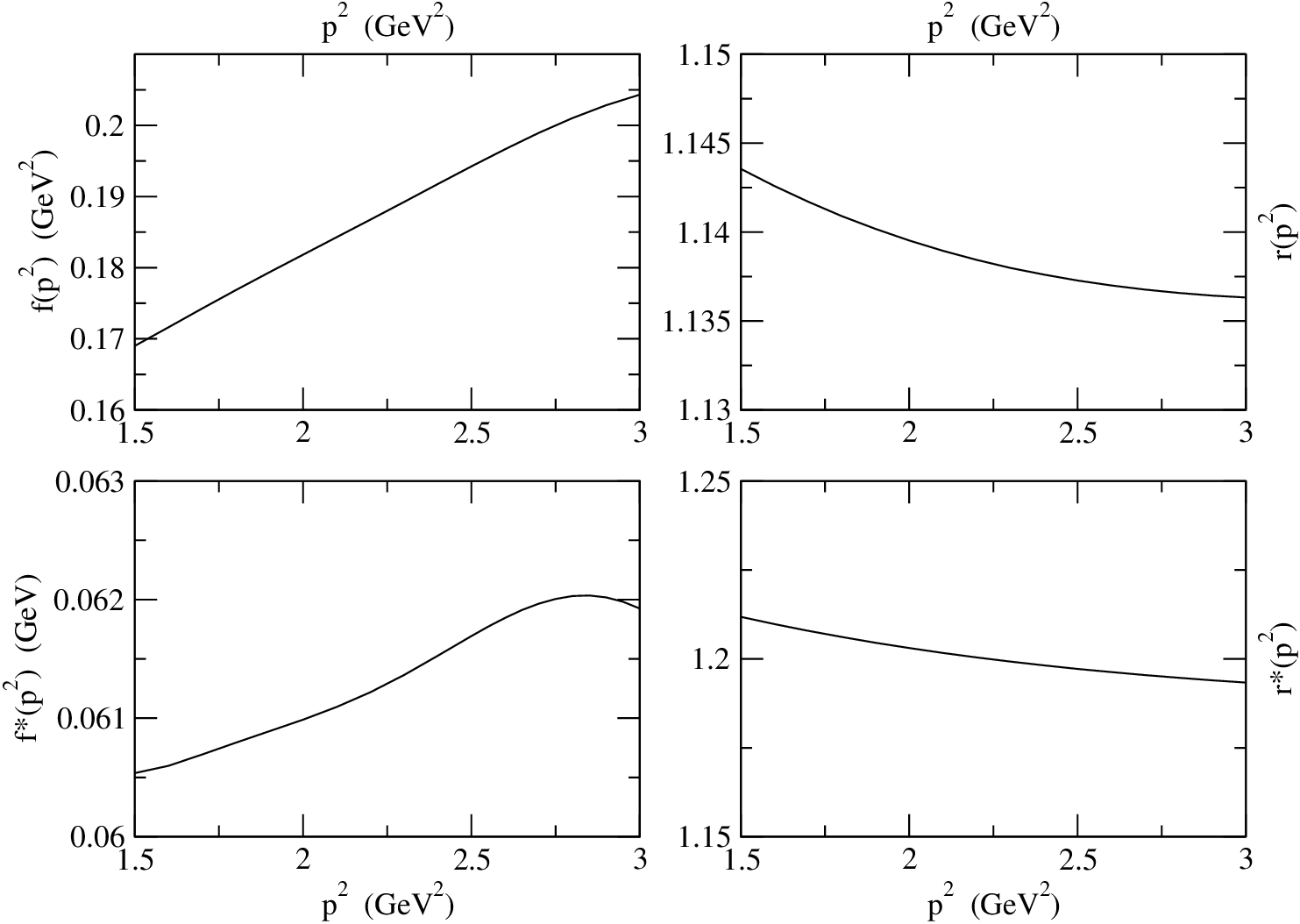}}
\caption{Plot of the mixing functions $f(p^{2}),r(p^{2})$ (eq. (\ref{compkg}%
)) and $f^{*}(p^{2}),r^{*}(p^{2})$ (eq. (\ref{complin})) for the
second propagator choice and with a vertex cut-off $\Lambda =1.5$ $GeV.$ }
\label{fig6}
\end{center}
\end{figure}

\end{document}